\documentclass[ammsymb,pre,aps,onecolumn,superscriptaddress,showpacs]{revtex4}
\usepackage{graphicx}
\usepackage{amsmath}
\usepackage{hyperref}
\usepackage{xcolor}

\begin{document}

\title{Contests in two fronts}

\author{A. de Miguel-Arribas}
\affiliation{Institute for Biocomputation and Physics of Complex Systems (BIFI), Edificio I + D, C/Mariano Esquillor s/n Campus R\'{i}o Ebro, University of Zaragoza, 50018 Zaragoza, Spain}

\author{J. Mor\'on-Vidal}
\affiliation{Department of Mathematics, Carlos III University of Madrid, 28911, Spain}

\author{L.M. Flor\'{i}a}
\affiliation{Institute for Biocomputation and Physics of Complex Systems (BIFI), Edificio I + D, C/Mariano Esquillor s/n Campus R\'{i}o Ebro, University of Zaragoza, 50018 Zaragoza, Spain}
\affiliation{Condensed Matter Department, Faculty of Sciences, University of Zaragoza, Pedro Cerbuna 12, 50009, Zaragoza, Spain}

\author{C. Gracia-L\'azaro}
\affiliation{Institute for Biocomputation and Physics of Complex Systems (BIFI), Edificio I + D, C/Mariano Esquillor s/n Campus R\'{i}o Ebro, University of Zaragoza, 50018 Zaragoza, Spain}

\author{L. Hern\'andez}
\affiliation{Laboratoire de Physique Théorique et Modélisation, UMR8089-CNRS, Université de Cergy-Pontoise, 2 Avenue Adolphe, Chauvin, F-95302, Cergy-Pontoise, Cedex, France}

\author{Y. Moreno}
\affiliation{Institute for Biocomputation and Physics of Complex Systems (BIFI), Edificio I + D, C/Mariano Esquillor s/n Campus R\'{i}o Ebro, University of Zaragoza, 50018 Zaragoza, Spain}
\affiliation{Theoretical Physics Department, Faculty of Sciences, University of Zaragoza, Pedro Cerbuna 12, 50009, Zaragoza, Spain}
\affiliation{Centai Institute, Turin, Italy}

\begin{abstract}
Within the framework of Game Theory, contests  study decision-making in those situations or conflicts when rewards depend on the relative rank between contenders rather than their absolute performance.
By relying on the formalism of Tullock  success functions, we propose a model where two contenders fight in a conflict on two fronts with different technology levels associated: a front with large resource demand and another with lower resource requirements. The parameter of the success function in each front determines the resource demand level. Furthermore, the redistribution or not of resources after a tie defines two different games. We solve the model analytically through the best-response map dynamics, finding a critical threshold for the ratio of the resources between contenders that determines the Nash Equilibrium basin and, consequently, the peace and fighting regimes. We also perform numerical simulations that corroborate and extend these findings. We hope this study will be of interest to areas as diverse as economic conflicts and geopolitics.

\end{abstract}
\maketitle

\vspace {-0.5cm}
\section{Introduction}
\label{sec:intro}

Contest Theory is a mathematical tool to model situations where two or more agents riskily compete, at a cost, for a prize \cite{tullock1967welfare,tullock1980efficient,krueger1974political,becker1983theory,corchon2018contest}. The strategic behavior in contests has attracted the attention of academia for many years \cite{dixit1987strategic,chowdhury2011generalized,vojnovic2015contest}, and has applications ranging from economics to conflict resolution and geopolitics. Actually, contests are studied in areas as diverse as labor economics, industrial organization, public economics, political science, rent-seeking, patent races, military combats, sports, or legal conflicts \cite{van2015theory,chowdhury2011generalized,connelly2014tournament}.

\vspace{0.3cm}

Formally, a contest is characterized by a set of agents, their respective possible efforts, a tentative payoff for each contestant (the prize), and a set of functions for the individual probabilities of obtaining the prize that takes the agents' efforts as parameters. The prize may, or not, be divisible, and contestants may or not have the same valuation of the prize \cite{corchon2018contest}.

\vspace{0.3cm}

A case of special interest is the contests in rent-seeking, which study those situations where there is no contribution of productivity nor added value \cite{tullock1980efficient,baye1993rigging}. Therefore, all the contenders' effort is devoted to winning the contest and so obtaining the whole payoff or the greatest possible share of it. This theoretical framework is applied to study issues such as elimination tournaments \cite{rosen1985prizes}, conflicts \cite{hirshleifer1989conflict,skaperdas1992cooperation}, political campaigns \cite{skaperdas1995modeling} or lobbying \cite{baye1993rigging,epstein2003lobbying}. 

\vspace{0.3cm}

In this regard, wars also constitute contests, where contenders compete for resources without adding productivity or value, being the appropriation of resources the main cause for war \cite{acemoglu2012dynamic,caselli2015geography,novta2016ethnic}, and therefore they are amenable to being theoretically studied as strategic tournaments 
\cite{dziubinski2021strategy,krainin2016war,levine2013conflict,piccione2007equilibrium,baliga2013bargaining,Baliga2012}. Similarly, in economic contests, resources allocation, and redistribution play also a key role in the strategic decision-making \cite{fahy2002role,connelly2014tournament}.

\vspace{0.3cm}

Despite a large amount of research on contest theory, most theoretical work is limited to one-front contests. Nevertheless, real-world competitions many times take place on two or more fronts. For example, a company fighting against a bigger one may be tempted to devote its resources (or some of them) to low-cost marketing instead of the costlier conventional one. This low-cost advertising, so-called \textit{guerrilla marketing} \cite{levinson1994guerrillaAdv,levinson1994guerrilla}, constitutes an active field of study \cite{baltes2008guerrilla,shakeel2011impact,nufer2013guerrilla,gokerik2018surprise}. Some examples of this guerrilla marketing are ambient advertising \cite{hutter2015unusual,gokerik2018surprise}, stealth marketing \cite{swanepoel2009virally}, word-of-mouth marketing \cite{chen2020psychology}, social media marketing \cite{alalwan2017social}, evangelism marketing \cite{anggraini2018understanding}, viral marketing \cite{reichstein2019decision}, or marketing buzz \cite{mohr2017managing}.
 
\vspace{0.3cm}

In this work, we focus on either armed or economic conflicts  
susceptible to being simultaneously fought on two front lines: one corresponding to a costly front (conventional war, costly marketing) and the other one to a low-cost front (guerrilla warfare/marketing). To that end, we rely on the formalism of Tullock's combats success functions by proposing two simultaneous fronts sustained by the same pair of contenders. Each of these fronts is characterized by a value of the parameter $\gamma$ of the Tullock function. The parameter $\gamma$ represents the technology associated with that front, i.e., the influence of the resources invested on the winning probability. The whole interaction constitutes a zero-sum game: the sum of the resources invested in both fronts makes up the total prize of the game or combat. That prize will go to the contender winning on both fronts if that is the case. Otherwise, i.e., if each contender wins in a front, we propose two scenarios, each constituting a different game. First, we consider those situations in which contenders recover their investments in case of a tie. This setup, hereafter the keeping resources game (KR), mimics those real-world conflicts where, after a tie, the previous \textit{status quo} is recovered, as mergers and acquisitions attempts in economics or, regarding army conflicts, abortable invasion temptations. The second setup, hereafter the redistributing resources game (RR), captures the cases where, after a tie, each contender gains all the resources invested in the front she won, like an open-ended long-term economic competition or war. 

\vspace{0.3cm}

In both setups, a contender will fight if her expected gains overcome her current resources. Then, peace takes place when no contender has the incentive to fight. Otherwise, the combat may repeat until i) one of the contenders wins on both fronts, taking all the resources, or ii) no contender has the incentive to fight.

\vspace{0.3cm}

We solve the system theoretically under the best-response dynamics, showing the existence, for both games, of two regimes regarding the ratio $r$ of contenders' resources: one with a Nash equilibrium and another without it. We also perform numerical simulations that confirm and extend the analytical results. In both games, the values of Tullock's technology parameters determine an $r$ threshold value, $r_{th}$, which points to the boundary between those regimes. This threshold demarcates the separation between war and peace: in the presence of a Nash equilibrium, the combat takes place and otherwise does not. Remarkably, in the KR game, peace takes place for high resource differences. Conversely, in the RR game, peace is reached for low differences.

\vspace{0.3cm}

The rest of the paper is organized as follows. The details of the model, together with combat functions and the best-response maps, are defined in Section \ref{sec:model}. In sections \ref{sec:KR} and \ref{sec:RR}, we study the KR and RR games, respectively. The repeated combats are studied in Section \ref{sec:repeated}. Finally, Section \ref{sec:conclusion} tries to summarize and contextualize the results together with prospective remarks.

\section{The model}
\label{sec:model}

Conflicts are not always amenable to reaching an agreement or peaceful solution, and ``win or lose'' scenarios (such as a war \cite{piccione2007equilibrium} or an economic contest \cite{hirshleifer1989conflict}) often emerge as the way out to their resolution. A useful, simple probabilistic description of the expected outcome of combat is provided by the formalism of contest success functions (CSF). A CSF \cite{skaperdas1996contest} is a function of the quantified efforts, or resources, invested by the contenders, that gives the probability of winning the contest. Though CSFs are in general defined for a number of contenders larger than two, we will restrict consideration to dyadic contests, and denote both contenders as {\bf 1} and {\bf 2}. 
\vspace{0.3cm}


Let $x$ be the resources of Contender \textbf{1} and $y$ those of Contender \textbf{2}. The CSF function called Tullock, for a positive parameter $\gamma$, meets the requirement that the winning probability $p$ of contender {\bf 1} is invariant under the re-scaling of both contenders' resources, i.e., for all $\lambda > 0$, $p(\lambda x, \lambda y)= p(x,y)$. Explicitly, the Tullock function:

\begin{equation}
\label{eq1.3}
p_{\gamma}(x, y) = \frac{x^{\gamma}}{x^{\gamma}+ y^{\gamma}} \;
\end{equation}
gives the winning probability of contender {\bf 1}. A basic assumption behind this result is that win and lose (from a contender perspective) are a mutually exclusive complete set of events, so that $p_{\gamma}(x, y) = 1-p_{\gamma}(y,x)$. 

\vspace{0.3cm}

Regarding the consequences of the contest outcome, one assumes that the winner's benefit is the sum $x+y$ of both resources, and the loser obtains nothing, zero benefits. From the, admittedly narrow, assumption of perfect rationality (i.e. the behavior is determined by the optimization of benefits), the decision to fight should be taken by a contender only if its expected gain after the contest is higher than its current resources.

\vspace{0.3cm}

In this regard, the parameter $\gamma$ of the Tullock CSF turns out to play a very important role, because when $\gamma>1$, it is easy to see that whenever $x>y$, the expected gain for the contender {\bf 1} after the combat, $p_{\gamma}(x,y) (x+y)>x$, and then the (richer) contender {\bf 1} has an incentive to fight, while if $\gamma<1$, the expected gain for the richer contender is lower than their resources before the combat, $p_{\gamma}(x,y) (x+y)<x$, and thus it is the poorer contender who should rationally decide to fight. 

\vspace{0.3cm}

Following the acutely descriptive terms introduced in \cite{dziubinski2021strategy}, we will call {\it rich-rewarding} a Tullock CSF with parameter $\gamma>1$, and {\it poor-rewarding} a Tullock CSF with $\gamma<1$. In this reference, \cite{dziubinski2021strategy}, where contests refer to events of ``real'' war among nations, a (highly costly) conventional war would be described by a rich-rewarding CSF, while guerrilla warfare would better be described by a poor-rewarding CSF Tullock function, which led the authors to refer to $\gamma$ as ``technology parameter'', and ponder its relevance to the expectations and chances for peaceful coexistence among nations or coalitions. Correspondingly, in economic contests, a rich-rewarding CSF corresponds to a competition in a conventional (costly) scenario and a poor-rewarding CSF to either a low-cost strategy or a guerrilla marketing scenario \cite{levinson1994guerrillaAdv}.

\vspace{0.3cm}

It is not hard to think of a conflict whose resolution is a war on several simultaneous fronts, each characterized by different Tullock parameters, where the ``rulers'' (decision-makers) of the two conflicting entities are faced with making a decision on the fraction of available resources that should be invested in each front.  We will consider here a war between two contenders which is conducted on two fronts, each one characterized by a different Tullock CSF. In the rich-rewarding front, the Tullock parameter is fixed to a value $\gamma_r >1$, while in the poor-rewarding front, the Tullock parameter is $\gamma_p<1$. Note that due to the scaling property of the Tullock function, the resources, $x$ and $y$, of the contenders can be rescaled to 1 and $r<1$, respectively, if we assume $x>y$, without loss of generality. After the rescaling, the contender {\bf 1} has resources $1$, of which a fraction $\alpha_1$ is invested in the rich-rewarding front (and then a fraction $1-\alpha_1$ is invested in the poor-rewarding front). The resources of the contender {\bf 2} are $r<1$, and its investment in the rich-rewarding front is $\alpha_2 r$ (and then its investment in the poor-rewarding front is $(1-\alpha_2) r$). Note that $0 \leq \alpha_1,\, \alpha_2 \leq 1$.

\vspace{0.3cm}

In the sequel, we will fix the values of the Tullock parameters, $\gamma_r >1$ (for the CSF of the rich-rewarding front) and $\gamma_p<1$ (poor-rewarding front), to some arbitrary values. We simplify a bit the notation for the winning probability of contender {\bf 1} at each front:

\begin{equation}
\label{eq1.1}
p (\alpha_1,\alpha_2)= \frac{\alpha_1^{\gamma_r}}{\alpha_1^{\gamma_r} + (\alpha_2 r)^{\gamma_r}}\;,\;\;\;\;\; q (\alpha_1,\alpha_2)= \frac{(1-\alpha_1)^{\gamma_p}}{(1-\alpha_1)^{\gamma_p} + ((1-\alpha_2) r)^{\gamma_p}}\;,
\end{equation}
and furthermore, we will simply write $p$ and $q$ whenever the arguments are unambiguous. The following relations concerning the partial derivatives of $p$ and $q$ are easily obtained:

\begin{equation}
\label{eq1.2}
\alpha_1 \frac{\partial p}{\partial \alpha_1} = - \alpha_2 \frac{\partial p}{\partial \alpha_2} = \gamma_r p(1-p)  \;,
\end{equation}

\begin{equation}
\label{eq1.3}
(1-\alpha_1) \frac{\partial q}{\partial \alpha_1} = - (1-\alpha_2) \frac{\partial q}{\partial \alpha_2} = - \gamma_p q(1-q) \;.
\end{equation}

\vspace{0.3cm}

We will consider the outcomes in both fronts as independent events, in the usual sense, so that e.g. the probability that contender {\bf 1} reaches victory on both fronts is the product $pq$. Also, whenever a contender wins on both fronts she obtains resources $1+r$, and her opponent receives zero resources. In the event of a tie, in which each contender reaches victory in only one front and is defeated in the other, we will consider two different rules that define two different games: 

\vspace{0.3cm}

\begin{itemize}
\item[{\bf KR}.] 
In the KR (keeping resources) game, if none of the contenders wins in both fronts, each one keeps their initial resources after the tie.
\item [{\bf RR}.] In the RR (redistributing resources) game, each contender receives the sum of the resources invested in the front where she has reached victory.
\end{itemize}

\vspace{0.3cm}

We denote by $u_i (\alpha_1,\alpha_2)$ ($i=1,2$), the expected gain of the contender {\bf{\it i}}. 

\vspace{0.3cm}

We call $\beta_1$ the best-response map of contender {\bf 1}, defined as follows: 

\begin{equation}
\label{eq2.3}
u_1(\beta_1(s), s) = \max_{\alpha_1} u_1(\alpha_1, s) \;,
\end{equation}
i.e. $\beta_1(s)$ is the value of $\alpha_1$ that maximizes the expected gain of contender {\bf 1} for the fraction of resources $\alpha_2=s$ of contender {\bf 2} in the rich-rewarding front. Correspondingly, we denote by $\beta_2$ the best-response map of contender {\bf 2}:

\begin{equation}
\label{eq2.4}
u_2(t, \beta_2(t)) = \max_{\alpha_2} u_2(t, \alpha_2) \;.
\end{equation}

\vspace{0.3cm}

The best-response maps $\beta_i$ ($i=1,2$) are determined by the three parameters ($r$, $\gamma_r$, $\gamma_p$) that define each particular KR (or RR) game. One should not expect them to be smooth 1d functions of the unit interval, for the $\max$ operation might introduce, in general, non-analyticities (e.g., jump discontinuities).

\vspace{0.3cm}

An ordered pair $(\bar{\alpha}_1, \bar{\alpha}_2)$ is a Nash equilibrium if the following two conditions are satisfied:

\begin{equation}
\label{eq2.5}
\bar{\alpha}_1 = \beta_1(\bar{\alpha}_2) \;\;\;\; {\mbox{and}}\;\;\;\; \bar{\alpha}_2 = \beta_2(\bar{\alpha}_1) \;, 
\end{equation}
or, equivalently,

\begin{equation}
\label{eq2.6}
\bar{\alpha}_1 = \beta_1(\beta_2(\bar{\alpha}_1)) \;\;\;\; {\mbox{and}}\;\;\;\; \bar{\alpha}_2 = \beta_2(\beta_1(\bar{\alpha}_2)) \;.
\end{equation}

When the contenders' choices of resources' assignments are a Nash equilibrium, none of them has any incentive to deviate.

\section{Keeping resources when tying}
\label{sec:KR}

In this section, we study the KR game. In this game, i) if none of the contenders win on both fronts (i.e., a tie), both keep their initial resources, while ii) if one of them wins on both fronts, the final resources are $1+r$ for the winner and zero for the loser. Thus, the expected gain after the contest, $u_1$, for the contender {\bf 1} are:

\begin{equation}
\label{eq2.1}
u_1 (\alpha_1,\alpha_2) = pq (1+r) + (p(1-q) + q(1-p)) = pq(r-1) + p + q  \;,
\end{equation}
and the expected gain, $u_2$, for the contender {\bf 2} are, in turn:

\begin{equation}
\label{eq2.2}
u_2 (\alpha_1,\alpha_2) = 1+r - u_1 (\alpha_1,\alpha_2) = 1+r-pq(r-1) - p - q  \;,
\end{equation}
where we have omitted the dependence of $p$ and $q$ on $\alpha_1$ and $\alpha_2$, the fractions of resources invested in the rich-rewarding front. 

\vspace{0.3cm}

\subsection{The best-response maps}
\label{subsec:KR_brm}

First, we now obtain the main features of the best-response map $\beta_1(s)$ of the contender {\bf 1}, for which we focus attention on its expected gain $u_1$ (equation (\ref{eq2.1})) as a function of its first argument $\alpha_1$, for fixed arbitrary values of its second argument $\alpha_2=s$. 

$$u_1(\alpha_1, s) = (r-1) \frac{\alpha_1^{\gamma_r}}{\left(\alpha_1^{\gamma_r}+(sr)^{\gamma_r}\right)} \frac{(1- \alpha_1)^{\gamma_p}}{\left((1- \alpha_1)^{\gamma_p} +( (1- s)r)^{\gamma_p}\right)} +\frac{\alpha_1^{\gamma_r}}{\left(\alpha_1^{\gamma_r}+(sr)^{\gamma_r}\right)} +\frac{(1- \alpha_1)^{\gamma_p}}{\left((1- \alpha_1)^{\gamma_p} +( (1- s)r)^{\gamma_p}\right)}$$

\vspace{0.3cm}

For $s=0$, one has
\begin{equation}
\label{eq2.3}
u_1 (\alpha_1,0) = 1 + r \frac{(1-\alpha_1)^{\gamma_p}}{(1-\alpha_1)^{\gamma_p}+ r^{\gamma_p}}  \;;
\end{equation}
this is a monotone decreasing function, thus taking its maximum value at the origin. However $\alpha_1=0$ and $s=0$ corresponds to the situation in which none of the contenders invests in the rich-rewarding front, and then the expected gain for the contender {\bf 1} is $u_1(0,0) = (1+r) (1+r^{\gamma_p})^{-1}$, i.e. the product of the total resources and the probability of victory in the poor-rewarding front. This is lower than the limit of expression (\ref{eq2.3}) when $\alpha_1 \rightarrow 0^+$:
\begin{equation}
\label{eq2.4}
u_1 (0^+,0) = 1 + r \frac{1}{1+ r^{\gamma_p}} > \frac{1+r}{1+r^{\gamma_p}}  = u_1(0,0) \;.
\end{equation}
In other words, the best response of contender {\bf 1} to $s=0$ is to invest as small as possible a positive quantity, say $\beta_1(0) = 0^+$.

\vspace{0.3cm}

Next, let us consider small positive values of $s$. For values of $\alpha_1$ such that $0<s\ll \alpha_1<1$, the expected gain $u_1(\alpha_1,s)$ is essentially given by $u_1 (\alpha_1,0)$, equation (\ref{eq2.3}):
\begin{equation}
\label{eq2.5}
u_1 (\alpha_1,s) \simeq 1 + r \frac{(1-\alpha_1)^{\gamma_p}}{(1-\alpha_1)^{\gamma_p}+ r^{\gamma_p}} \;\;\;\; {\mbox{for}}\;\; \alpha_1\gg s>0 \;.
\end{equation}

\begin{figure}
  \centering
  \includegraphics[width=0.75\linewidth]{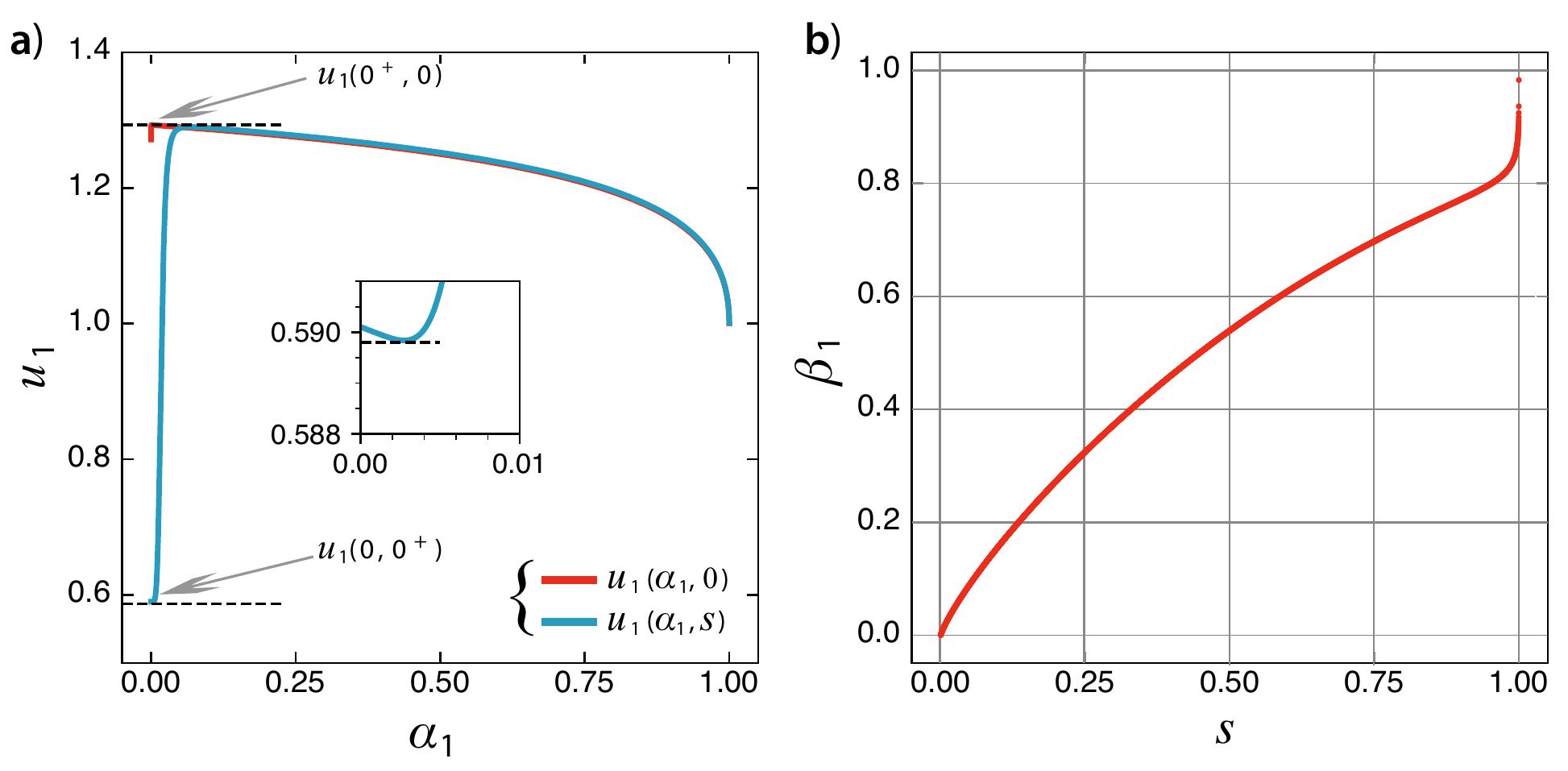}
  \caption{KR game with parameters $r=0.5$, $\gamma_r= 5$, and $\gamma_p=0.5$. Panel \textbf{a} (left) shows the graphs of the contender {\bf 1} expected gain, $u_1$, as a function of its investment fraction $\alpha_1$ in the rich-rewarding front (RRF), for $s=0$ (red) and $s=0.035$ (purple), where $s$ is the contender {\bf 2} invested fraction of resources in RRF. The local minimum of $u_1(\alpha_1, s=0.035)$ is shown in the inset. Panel \textbf{b} (right) shows the best-response map $\beta_1(s)$. See text for details.}
  \label{fig1}
\end{figure}

However, for lower values of $\alpha_1$, $u_1(\alpha_1,s)$ differs significantly from (\ref{eq2.5}). In particular, for $\alpha_1=0$ the expected gain is
\begin{equation}
\label{eq2.6}
u_1 (0,s) = \frac{1}{1+ ((1-s) r)^{\gamma_p}}  \;,
\end{equation}
and $u_1 (\alpha_1,s)$ is a decreasing function at the origin:
\begin{equation}
\label{eq2.7}
\left. \frac{\partial u_1}{\partial \alpha_1}\right|_{\alpha_1=0} \equiv u'_1 (0,s) = -\frac{\gamma_p ((1-s)r)^{\gamma_p}}{(1+ ((1-s) r)^{\gamma_p})^2}  \;.
\end{equation}
It can be shown that when $\alpha_1$ increases from zero, the function $u_1 (\alpha_1,s)$ shows a local minimum, followed by a local maximum before it approaches (\ref{eq2.5}). Both, the locations of the minimum and the maximum tend to zero as $s \rightarrow 0$; in this limit, the value of $u_1$ at the maximum converges to $u_1 (0^+,0)$, see equation (\ref{eq2.4}), while its value at the minimum tends to $u_1 (0,0^+) = \frac{1}{1+  r^{\gamma_p}} < u_1 (0^+,0)$. Thus the location of the local maximum gives the value of the best-response map $\beta_1(s)$, for (\ref{eq2.5}) is monotone decreasing. To illustrate this, Panel \textbf{a} of Figure \ref{fig1} displays, for an exemplifying choice $r=0.5$, $\gamma_r= 5$, $\gamma_p=0.5$, the numerical results corresponding to the expected gain $u_1(\alpha_1,s)$ of Contender \textbf{1}, as a function of the fraction $\alpha_1$ that she invested in the rich-rewarding front, for two fixed values ($0,\;0.035$) of the Contender \textbf{2} invested fraction $s$ in the reach-rewarding front. Note the non-monotonous behavior for $s>0$ (purple line). The inset highlights the local minimum and maximum for $s>0$.

\vspace{0.3cm}

The conclusion of the analysis for small values of $s \ll 1$ is that the best-response map $\beta_1(s)$ is a well-behaved monotone increasing function in this region of $s$ values. For generic, not too small values of $s$, the qualitative features of the function $u_1 (\alpha_1,s)$ remain the same: it shows a negative slope at the origin, a local minimum followed by a local maximum, and a divergent ($- \infty$) slope at $\alpha_1=1$ (so that it is ensured that $\beta_1(s)<1$ for all values of $s$). However, its maximum value is no longer guaranteed to occur at its local maximum, for it can perfectly occur at the origin, as the position of the local maximum increases with $s$ (and then the value of $u_1$ decreases there) while the value of $u_1 (0,s)$ increases, see equation (\ref{eq2.6}).
In other words, the continuity of $\beta_1(s)$ is not guaranteed. To explore this shape, we have computed the best-response map $\beta_1(s)$ for the specific set of values $r=0.5$, $\gamma_r= 5$, and $\gamma_p=0.5$. Panel \textbf{b} of Figure \ref{fig1} displays the numerical results for the best response of Contender \textbf{1} to Contender \textbf{2}'s rich-rewarding-front investment ratio $s$. As shown, for these values of the contenders' resources ratio and Tullock function parameters,
$\beta_1(s)$ is a well-behaved monotone increasing function in the whole range $0<s<1$.

\vspace{0.3cm}

To obtain the main features of the best-response map $\beta_2(t)$ of the contender {\bf 2}, we analyze its expected gain $u_2$ as a function of its second variable $\alpha_2$ for fixed values of $\alpha_1=t$.

$$u_2(t,\alpha_2) =1+r- (r-1) \frac{t^{\gamma_r}}{\left(t^{\gamma_r}+(\alpha_2 r)^{\gamma_r}\right)} \frac{(1- t)^{\gamma_p}}{\left((1- t)^{\gamma_p} +( (1- \alpha_2)r)^{\gamma_p}\right)} -\frac{t^{\gamma_r}}{\left(t^{\gamma_r}+(\alpha_2 r)^{\gamma_r}\right)} -\frac{(1- t)^{\gamma_p}}{\left((1- t)^{\gamma_p} +( (1- \alpha_2)r)^{\gamma_p}\right)}$$

For $t=0$, $u_2(0,\alpha_2)$ is a monotone decreasing function of $\alpha_2$:
\begin{equation}
\label{eq2.8}
u_2 (0,\alpha_2) = 1 + r -\frac{1}{1+((1-\alpha_2) r)^{\gamma_p}}  \;.
\end{equation}
However, in a similar way as we discussed above for the function $u_1(\alpha_1,0)$, due to the discontinuity of $u_2 (0,\alpha_2)$ at the origin, i.e.
\begin{equation}
\label{eq2.9}
u_2 (0,0^+) \equiv \lim_{\alpha_2 \rightarrow 0} u_2 (0,\alpha_2) = 1 + r -\frac{1}{1+ r^{\gamma_p}} > \frac{(1+r) r^{\gamma_p}}{1+r^{\gamma_p}}  = u_2(0,0) \;,
\end{equation}
the best response of contender {\bf 2} to $t=0$ is to invest as small as possible a positive quantity, say $\beta_2(0) = 0^+$.

\vspace{0.3cm}

The analysis of $u_2(t,\alpha_2)$ for small positive values of $t$ is similar to that of $u_1(\alpha_1,s)$ for small positive values of $s$, and leads to analogous conclusions, i.e. the function $u_2(t,\alpha_2)$ shows a local minimum followed by a local maximum before approaching expression (\ref{eq2.8}). The location of this local maximum gives the best-response map $\beta_2(t)$, and thus this map is a well-behaved monotone increasing function for small positive values of $t$. 

\vspace{0.3cm}

\begin{figure}
  \centering
  \includegraphics[width=0.75\linewidth]{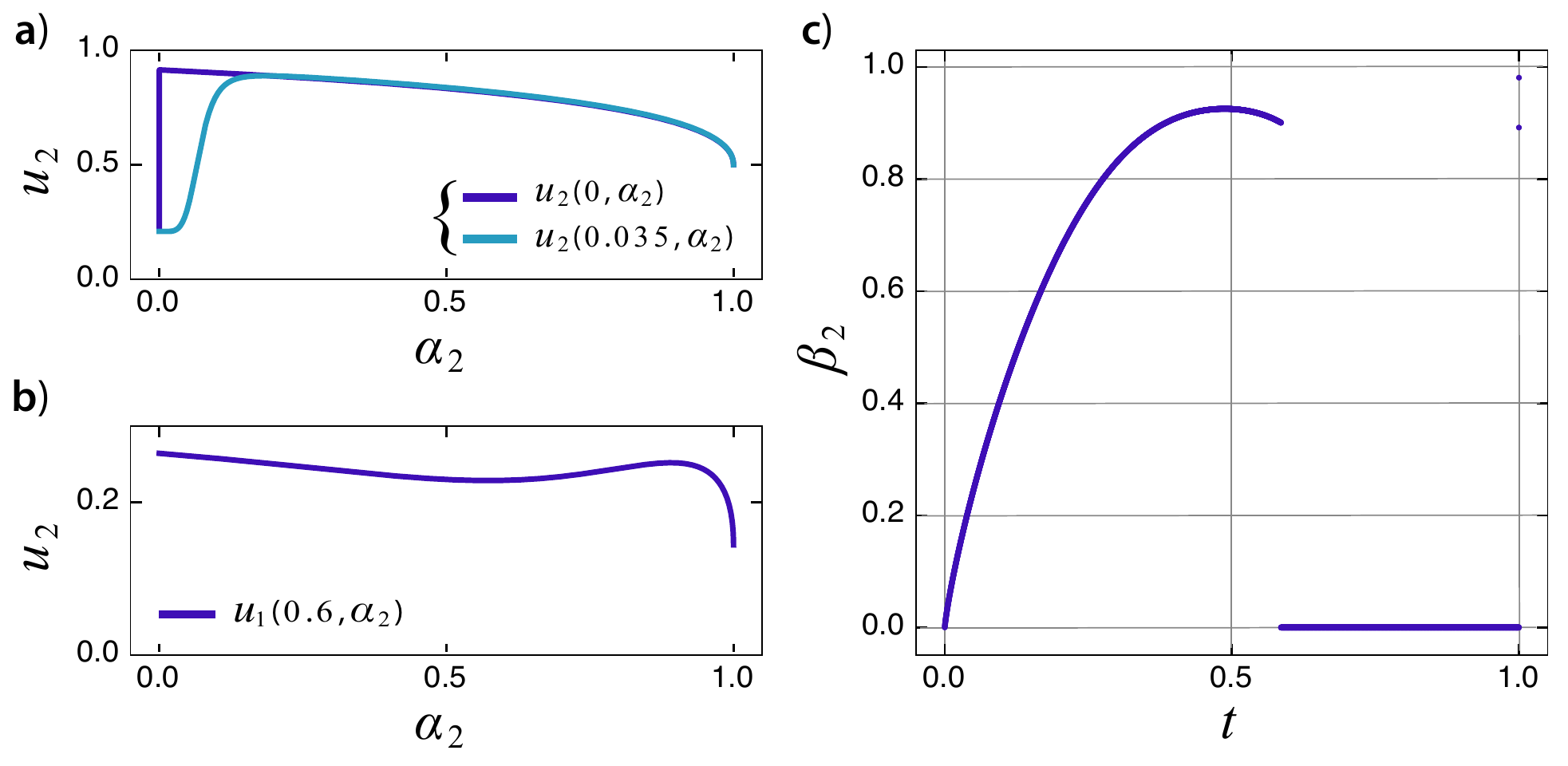}
  \caption{
  KR game with parameters $r=0.5$, $\gamma_r= 5$, and $\gamma_p=0.5$. Left panels (\textbf{a} and \textbf{b}) show the graphs of the contender {\bf 2} expected gain, $u_2$, as a function of its investment fraction $\alpha_2$ in the rich-rewarding front (RRF), for $t=0$ (Panel \textbf{a}, blue line), $t=0.035$ (Panel \textbf{a}, purple), and $t=0.6$ (Panel \textbf{b}), where $t$ is the contender {\bf 1} invested fraction of resources in RRF. Panel \textbf{c} shows the best-response map $\beta_2(t)$. See text for details.}
  \label{fig2}
\end{figure}

These qualitative features of $u_2(t,\alpha_2)$ remain unaltered for generic, not too small, values of $t$. Also, its maximum cannot occur at $\alpha_2=1$ because its slope there diverges to $- \infty$. And again, there is no guarantee that the best-response map is given by the location of the local maximum of $u_2(t,\alpha_2)$, for $u_2(t,0)$ keeps growing with increasing values of $t$, so that an eventual jump discontinuity where $\beta_2(t)$ drops to zero may occur. As for Contender \textbf{1}, we have numerically explored the expected gain and best response of Contender \textbf{2}. Panels \textbf{a} (top) and \textbf{b} (bottom) of Figure \ref{fig2} display, for $r=0.5$, $\gamma_r= 5$, $\gamma_p=0.5$, the expected gain $u_2(t,\alpha_2)$ of Contender \textbf{2} versus the fraction $\alpha_2$ she invested in the rich-rewarding front, for three fixed values of the relative investment of Contender \textbf{1} in the reach-rewarding front. As predicted, the numerical results confirm the non-monotonous behavior for $t>0$. Panel \textbf{c} (right) displays the best-response map $\beta_2(t)$ for Contender \textbf{2}, showing the aforementioned discontinuity, $\beta_2(t)$ dropping to zero at $t\simeq0.585$ for the chosen values ($r=0.5$, $\gamma_r= 5$, $\gamma_p=0.5$).

\vspace{0.3cm}

\subsection{The Nash equilibrium}
\label{subsec:KR_ne}

The previous characterization of the best-response maps, $\beta_1(s)$ and $\beta_2(t)$, leads to the conclusion that a Nash equilibrium $(\bar{\alpha}_1, \bar{\alpha}_2)$  of a KR game must be an interior point of the unit square, i.e. $0< \bar{\alpha}_1,\, \bar{\alpha}_2 <1$. Indeed, on one hand, $\beta_1(s) \neq 1$ for all $s$, and $\beta_2(t) \neq 1$, for all $t$. On the other hand, $\beta_i(0)$ ($i=1,2$) is a small positive quantity and then it is ensured that $\beta_j(\beta_i(0))$ is a positive quantity. The important consequence is that any Nash equilibrium of a KR game must solve for the system of equations:
\begin{equation}
\label{eq2.10}
\frac{\partial u_1 (\alpha_1,\alpha_2)}{\partial \alpha_1} = 0 \;, \;\;\;\; \frac{\partial u_2 (\alpha_1,\alpha_2)}{\partial \alpha_2} = 0 \;.
\end{equation}
 
\vspace{0.3cm}

Using the equalities (\ref{eq1.2}) and (\ref{eq1.3}), the system (\ref{eq2.10}) is written as:
\begin{equation}
\label{eq2.11}
\frac{\alpha_1}{1- \alpha_1} = f(\alpha_1, \alpha_2) \;, \;\;\;\; \frac{\alpha_2}{1- \alpha_2} = f(\alpha_1, \alpha_2) \;,
\end{equation}
where $f(\alpha_1, \alpha_2)$ is the following function:
\begin{equation}
\label{eq2.12}
 f(\alpha_1, \alpha_2) = \frac{\gamma_r p(1-p)}{\gamma_p q(1-q)} \; \frac{1+(r-1) q}{1+(r-1) p} \;.
\end{equation}

\vspace{0.3cm}

First, one sees that $\bar{\alpha_1} = \bar{\alpha_2} \equiv \bar{\alpha}$. Then, due to the scaling property of the Tullock functions, (\ref{eq2.11}) becomes a simple linear equation 
\begin{equation}
\label{eq2.13}
\frac{\bar{\alpha}}{1- \bar{\alpha}}  = \bar{f} \equiv \frac{\gamma_r (1+r^{\gamma_p})(1+r^{1-\gamma_p})}{\gamma_p (1+r^{\gamma_r})(1+r^{1-\gamma_r})}  \;,
\end{equation}
with a unique solution (for fixed $r$, $\gamma_r$ and $\gamma_p$ values) given by
\begin{equation}
\label{eq2.14}
\bar{\alpha} = \frac{\bar{f}}{1 + \bar{f}} \;.
\end{equation}

\vspace{0.3cm}

\begin{figure}
  \centering
  \includegraphics[width=0.75\linewidth]{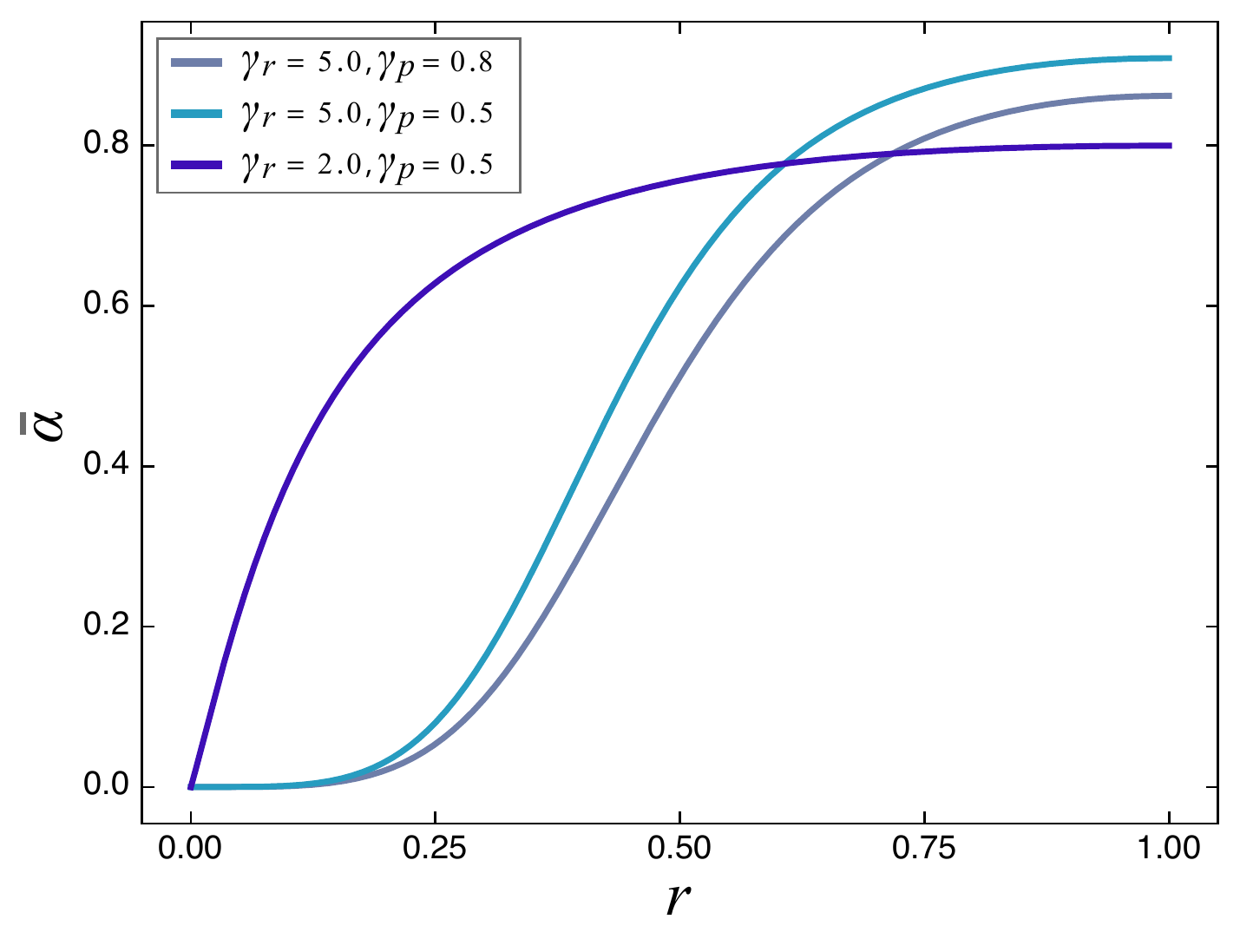}
  \caption{KR game. Graph of the function $\bar{\alpha}(r)$ for three different pairs of values (shown in legend) of the technology parameters $(\gamma_r, \gamma_p)$. The point $(\alpha_1,\alpha_2)=( \bar{\alpha},\bar{\alpha})$ corresponds to the local maxima of the expected gain $u_1(\alpha_1, \bar{\alpha})$, $u_2(\bar{\alpha}, \alpha_2),$ for both contenders, where $\alpha_1$ (resp., $\alpha_2$) is the fraction invested by Contender \textbf{1} (resp., \textbf{2}) in the rich-rewarding front. See the text for further details.}
  \label{fig3}
\end{figure}

In Figure \ref{fig3} we show the graph of the function $\bar{\alpha}(r)$ for three different pairs of values of the technology parameters $(\gamma_r, \gamma_p)$. Still, we should be aware that it is not guaranteed that for fixed values of $r$, $\gamma_r$, and $\gamma_p$, the pair ($\bar{\alpha}, \bar{\alpha}$) is a Nash equilibrium. So far, we have only shown that $\bar{\alpha}$ is a local maximum of $u_1(\alpha_1, \bar{\alpha})$ and a local maximum of $u_2(\bar{\alpha}, \alpha_2)$; because any Nash equilibrium of a KR game must be an interior point, this is a necessary condition, but not a sufficient one. 

\vspace{0.3cm}

The solution ($\bar{\alpha}, \bar{\alpha}$) of the system of equations (\ref{eq2.10}) is a Nash equilibrium of the KR game if the following conditions are satisfied:
\begin{itemize}
\item[C1.-] $ \bar{\alpha}$ is a global maximum of $u_1(\alpha_1, \bar{\alpha})$, i.e.:
\begin{equation}
\label{eq2.15}
u_1(\bar{\alpha}, \bar{\alpha}) > u_1(0, \bar{\alpha}) \;, \;\;{\mbox{and}}\;\;\; u_1(\bar{\alpha}, \bar{\alpha}) > u_1(1, \bar{\alpha}) \;.
\end{equation}

\item[C2.-] $ \bar{\alpha}$ is a global maximum of $u_2( \bar{\alpha}, \alpha_2)$, i.e.:
\begin{equation}
\label{eq2.16}
u_2(\bar{\alpha}, \bar{\alpha}) > u_2( \bar{\alpha}, 0) \;, \;\;{\mbox{and}}\;\;\; u_2(\bar{\alpha}, \bar{\alpha}) > u_2(\bar{\alpha},1) \;.
\end{equation}

\end{itemize}

\vspace{0.3cm}

\vspace{0.3cm}

It is straightforward to check that $$u_1(\bar{\alpha}, \bar{\alpha}) = \frac{1}{1+r^{\gamma_r}} + \frac{1}{1+r^{\gamma_p}} +(r-1) \frac{1}{1+r^{\gamma_r}} \;\frac{1}{1+r^{\gamma_p}} > 1 \;,$$
while $$u_1(0, \bar{\alpha}) = q(0, \bar{\alpha}) <1\;, \;\;{\mbox{and}}\;\;\; u_1(1, \bar{\alpha})=p(1,\bar{\alpha}) <1\;,$$
and one concludes that the conditions C1 are satisfied for all values of the game parameters $r$, $\gamma_r$ and $\gamma_p$.

\vspace{0.3cm}

\begin{figure}
  \centering
  \includegraphics[width=0.75\linewidth]{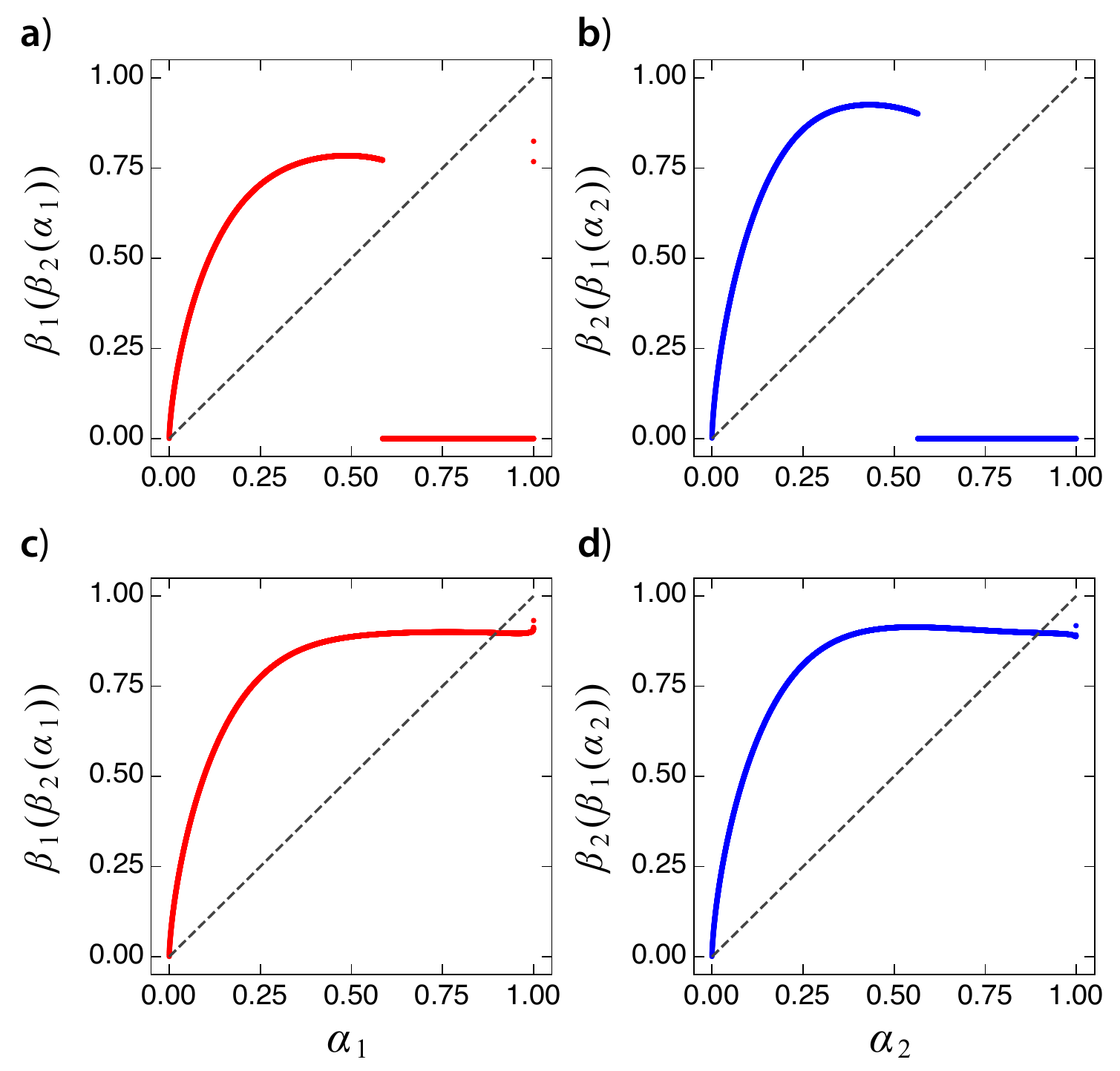}
  \caption{KR game with $\gamma_r=5$ and $\gamma_p=0.5$. Plots of the composition of players' best-response maps for $r=0.5$ (top panels, \textbf{a} and \textbf{b}) and $r=0.85$ (bottom panels, \textbf{c} and \textbf{d}). Left panels (\textbf{a} and \textbf{c}) show $\beta_1(\beta_2( \alpha_1))$, while $\beta_2(\beta_1( \alpha_2))$ is shown in right panels (\textbf{b} and \textbf{d}). The main diagonal (in dashed black) is plotted to visualize the existence for $r=0.85$ of a Nash equilibrium, and its absence for $r=0.5$.}
  \label{fig4}
\end{figure}
\vspace{0.3cm}

On the contrary, one can easily find values of the game parameters where the conditions C2 do not hold, as well as other values for which they do. As an illustrative example, we show in Figure \ref{fig4} the graphs of the best-response maps for the Tullock parameters $\gamma_r=5$ and $\gamma_p=0.5$, relative to $r=0.5$ (top panels, \textbf{a} and \textbf{b}) and $r=0.85$ (bottom, \textbf{c} and \textbf{d}), left panels (\textbf{a} and \textbf{c}) corresponding to $\beta_1(\beta_2( \alpha_1))$ and right ones (\textbf{b} and \textbf{d}) to $\beta_2(\beta_1( \alpha_2))$. An inner intersection of the curve with the black main diagonal indicates the existence of a Nash equilibrium. As shown in this example $(\gamma_r=5 , \gamma_p=0.5)$, for $r=0.85$, there is a Nash equilibrium, while for $r=0.5$, there is not. Our extensive exploration of the ($\gamma_r, \gamma_p$) plane strongly suggests that there is a threshold value $r_{th} (\gamma_r, \gamma_p)$, that depends on the Tullock parameters, such that for $r > r_{th}$ both conditions C2 are satisfied. In this case, the corresponding KR game has a Nash equilibrium, where both contenders invest a fraction $\bar{\alpha}(r, \gamma_r, \gamma_p)$ of their resources in the rich-rewarding front.

\vspace{0.3cm}

The existence of a Nash equilibrium given by the pair ($\bar{\alpha}, \bar{\alpha}$) for large enough values of the parameter $r$ can be proved by a continuation argument from the ``equal resources'' limit $r=1$, where one can directly check that the conditions C2 hold. Indeed, in this limit $\bar{f}= \gamma_r/\gamma_p$, and then $$\bar{\alpha}(r=1)= \frac{\gamma_r}{\gamma_r + \gamma_p} < 1\;, \;\;{\mbox{and}}\;\;\; u_2(\bar{\alpha}, \bar{\alpha})=1\;,$$
while $$u_2( \bar{\alpha}, 0)= \left( 1+ \left(\frac{\gamma_p}{\gamma_r + \gamma_p}\right)^{\gamma_p} \right)^{-1} < 1\;, \;\;{\mbox{and}}\;\;\;  u_2(\bar{\alpha},1)= \left( 1+ \left(\frac{\gamma_r}{\gamma_r + \gamma_p}\right)^{\gamma_p} \right)^{-1} < 1 \;,$$
and thus conditions C2 are satisfied in the equal resources limit.

\vspace{0.3cm}

\begin{figure}
  \centering
  \includegraphics[width=0.75\linewidth]{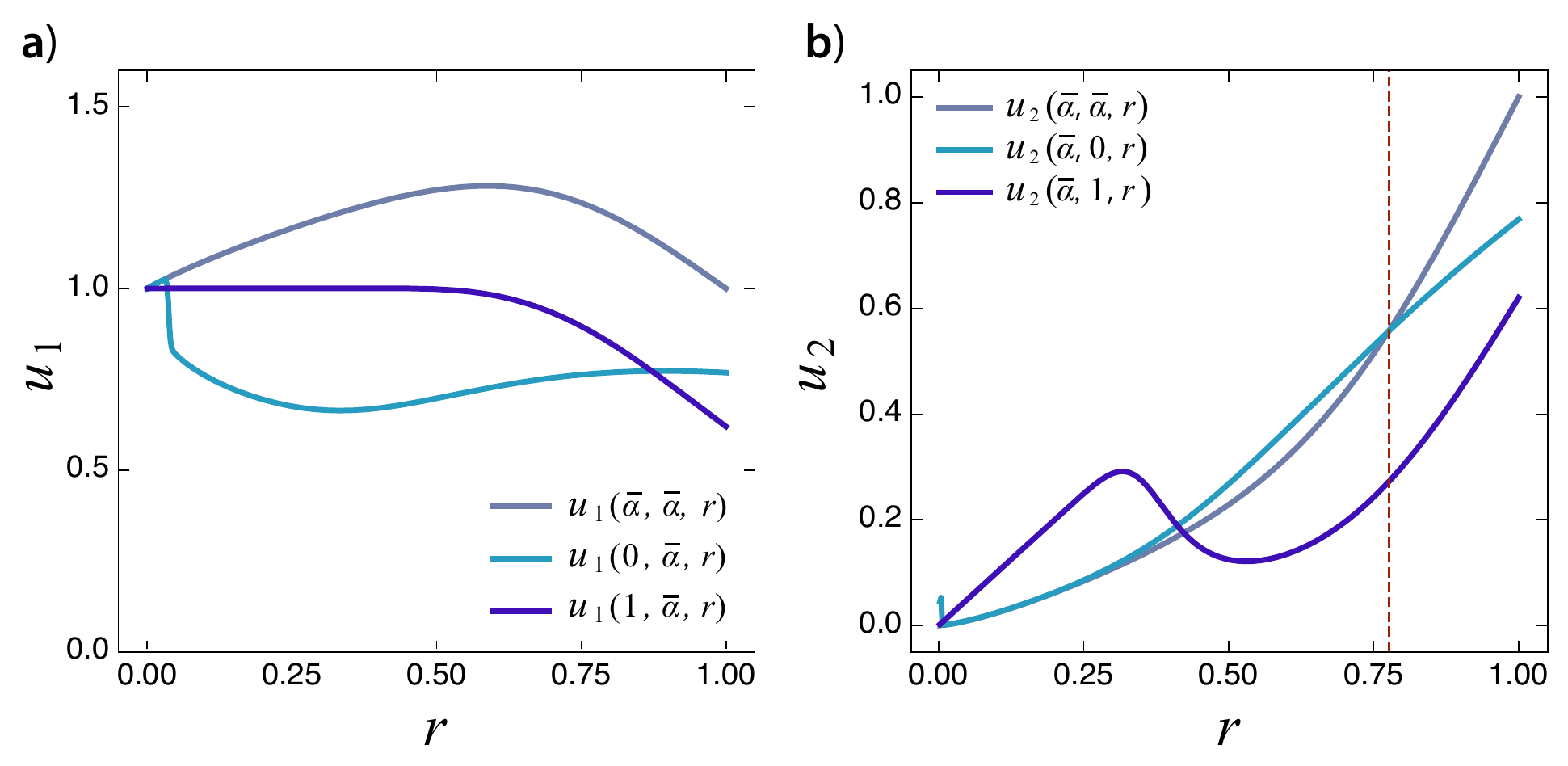}
  \caption{Illustration of C1 conditions (Panel \textbf{a}) and C2 conditions (Panel \textbf{b}) for RR game. Panel \textbf{a} depicts the expected gain $u_1(r)$ evaluated at $(\alpha_1,\alpha_2)=(\bar{\alpha},\bar{\alpha})$ (blue curve), $(0,\bar{\alpha})$ (purple) and $(1,\bar{\alpha})$ (red). Similarly, Panel \textbf{b} depicts the expected gain $u_2(r)$ evaluated at $(\alpha_1,\alpha_2)=(\bar{\alpha},\bar{\alpha})$ (blue curve), $(\bar{\alpha},0)$ (purple) and $(\bar{\alpha},1)$ (red). The vertical dashed line marks at $r=0.77635$ the point where conditions C1 and C2 start to be simultaneously satisfied.}
  \label{fig5}
\end{figure}
\vspace{0.3cm}

In Figure \ref{fig5}, Panel \textbf{a} shows, for $\gamma_r=5$ and $\gamma_p=0.5$, the graph of $u_1(\bar{\alpha}, \bar{\alpha})$ as a function of $r$, along with $u_1(0, \bar{\alpha})$ and $u_1(1, \bar{\alpha})$, to illustrate the conditions C1. Panel \textbf{b} displays $u_2(\bar{\alpha}, \bar{\alpha})$, $u_2(\bar{\alpha}, 0)$, and $u_2(\bar{\alpha}, 1)$, showing that conditions C2 are only satisfied simultaneously for $r>0.77635$ (dashed vertical line).

\vspace{0.3cm}

\section{Redistributing resources when tying}
\label{sec:RR}

In this section, we analyze the RR game, in which ties are followed by a redistribution of resources among the contenders that depend on their investments on each front. Specifically, each contender collects the sum of the investments employed in the front where she reached victory. Thus the expected gain is:

\begin{equation}
\label{eq3.1}
u_1 = (\alpha_1 + \alpha_2 r) p + \left( (1-\alpha_1) + (1-\alpha_2)r \right) q \;,
\end{equation}

\begin{equation}
\label{eq3.2}
u_2 = (\alpha_1 + \alpha_2 r) (1-p) + \left( (1-\alpha_1) + (1-\alpha_2)r \right) (1-q) \;,
\end{equation}
where the winning probabilities, $p$ and $q$, of contender {\bf 1} in each front are given by equation (\ref{eq1.1}).

\vspace{0.3cm}

\subsection{The best-response maps}
\label{subsec:RR_brm}

First, let us consider the expected gain $u_1$ of contender {\bf 1} as a function of $\alpha_1$, for a fixed value of $\alpha_2=s$, 
\begin{equation}
\label{eq3.3}
u_1(\alpha_1,s) = (\alpha_1 + s r) \frac{\alpha_1^{\gamma_r}}{\alpha_1^{\gamma_r}+ (sr)^{\gamma_r}} + \left( (1-\alpha_1) + (1-s)r \right) \frac{(1-\alpha_1)^{\gamma_p}}{(1-\alpha_1)^{\gamma_p}+ ((1-s)r)^{\gamma_p}} \;.
\end{equation}

For $s=0$, we have
\begin{equation} 
\label{eq3.4}
u_1(\alpha_1,0) = \alpha_1 + (1-\alpha_1+ r) \frac{(1-\alpha_1)^{\gamma_p}}{(1-\alpha_1)^{\gamma_p}+ r^{\gamma_p}} \;.
\end{equation}

Note that, contrary to the situation in the KR game, analyzed in the previous section \ref{subsec:KR_brm}, this is a continuous function at the origin:
\begin{equation} 
\label{eq3.5}
u_1(0^+,0) = u_1(0,0) =  \frac{1+r}{1+ r^{\gamma_p}} <1 \;.
\end{equation}

As $u_1(1,0) = 1$, it is plain that $\beta_1(0) \neq 0$. Furthermore, the first derivative of $u_1(\alpha_1,0)$, given by
\begin{equation}
\label{eq3.6}
u'_1 (\alpha_1,0) =  \frac{r^{\gamma_p}}{(1-\alpha_1)^{\gamma_p} + r^{\gamma_p}} \left( 1- \left( 1+ \frac{r}{1-\alpha_1} \right) \frac{\gamma_p (1-\alpha_1)^{\gamma_p}}{(1-\alpha_1)^{\gamma_p} + r^{\gamma_p}} \right) \;,
\end{equation}
is positive at the origin,
\begin{equation}
\label{eq3.7}
u'_1 (0,0) =  \frac{r^{\gamma_p}}{1 + r^{\gamma_p}} \left( 1-  \frac{\gamma_p (1+r)}{1 + r^{\gamma_p}} \right)>0 \;,
\end{equation}
and diverges to $-\infty$ as $\alpha_1\rightarrow1$, as 
\begin{equation}
\label{eq3.8}
u'_1 (1^-,0) \sim  (1-\alpha_1)^{\gamma_p - 1} \;;
\end{equation}
thus $\beta_1(0) \neq 1$, and $\beta_1(0)$ must be an interior point $0<\alpha_1^* < 1$. Then one concludes that $\alpha_1^*$ must solve for the equation
\begin{equation}
\label{eq3.9}
u'_1 (\alpha_1,0) = 0 \;.
\end{equation}

From (\ref{eq3.6}), with the change of variable $z\equiv r/(1 - \alpha_1)$, we can simply write (\ref{eq3.9}) in terms of $z$ as
\begin{equation}
\label{eq3.10}
\gamma_p (1+z) = 1 + z^{\gamma_p} \;.
\end{equation}
Note that as $s=0$, the variable $z$ is no other thing than the ratio of the resources invested by the contenders in the poor-rewarding front. This is so since $(1-\alpha_2)y/[(1-\alpha_1)x]$, and $\alpha_2=0$ in the RR game. Moreover, it is not difficult to realize that the equation (\ref{eq3.10}) has a unique positive solution, say $z^*$. Indeed, let us call $f(z)$ its LHS, and $g(z)$ its RHS; clearly $f(0)<g(0)$, while at very large values of $z \gg 1$, $z \gg z^{\gamma_p}$, so that $f(z)>g(z)$. Then, there exists at least a solution of (\ref{eq3.10}), and because $f(z)$ is linear and $g(z)$ is a convex function, the solution is unique.

\vspace{0.3cm}

It is worth remarking that $z^*$ is solely determined by the value of the Tullock parameter, $\gamma_p$, of the poor-rewarding front, and that $z^*(\gamma_p)$ is a monotone decreasing function of its argument. Thus, as $\gamma_p <1$, the value of $z^*$ is bounded below by  $z^*(1^-) \simeq 3.590175 > 1$, after carefully noticing that the correct limit when $\gamma_p \rightarrow 1$ of the equation (\ref{eq3.10}) is $1+z=z \ln z$.

\vspace{0.3cm}

The unique solution of the equation (\ref{eq3.9}), $\alpha_1^* = 1- r/z^*$ is clearly, due to (\ref{eq3.7}) and (\ref{eq3.8}), the maximum of $u_1(\alpha_1,0)$, and then,
\begin{equation}
\label{eq3.11}
\beta_1(0)= 1-\frac{r}{z^*}\;.
\end{equation}

\vspace{0.3cm}

For very small values of $s \ll 1$, if $\alpha_1$ is also small, $u_1(\alpha_1,s)$ differs qualitatively from $u_1(\alpha_1,0)$. Though for the expected gain $u_1$ we have that $u_1(0^+,0) = u_1(0,0)= u_1(0, 0^+)$, the first derivative of $u_1(\alpha_1, s)$ at $\alpha_1=0$, 
\begin{equation}
\label{eq3.12}
u_1'(0,s) = - \frac{1}{1 + ((1-s)r)^{\gamma_p}} \left(1 + \gamma_p(1 + (1-s)r) \frac{((1-s)r)^{\gamma_p}}{1 +((1-s)r)^{\gamma_p}} \right) < 0\;,
\end{equation}
converges, as $s \rightarrow 0$ to the limit
\begin{equation}
\label{eq3.13}
u_1'(0,0^+) = - \frac{1}{1 + r^{\gamma_p}} \left(1 + \gamma_p(1 + r) \frac{r^{\gamma_p}}{1 +r^{\gamma_p}} \right) < 0 < u'_1 (0,0) \;,
\end{equation}
where the last inequality comes from (\ref{eq3.7}). Then, $u_1(\alpha_1,s)$ is a decreasing function at the origin as soon as $s \neq 0$, showing a local minimum that detaches from 0 with increasing values of $s$. Also, it has a local maximum whose location $\alpha_1^*(s)$ is a smooth continuation of $\alpha_1^* = 1- r/z^*$, the maximum of $u_1(\alpha_1,0)$, because for $\alpha_1 \gg s$ both functions are uniformly very close each other. That local maximum is the value of the best-response map $\beta_1(s)$, for small values of $s$.

\vspace{0.3cm}

For larger values of $s$ the qualitative features of $u_1(\alpha_1,s)$ remain the same. The location of its local maximum $\alpha_1^*(s)$ increases with $s$, and, as $u_1(0,s)$ is a decreasing function of $s$, its value remains lower than $u_1(\alpha_1^*(s),s)$. Then $\beta_1(s)=\alpha_1^*(s)$ increases smoothly, and approaches the value 1, as $s\rightarrow 1$, with no jump discontinuities.

\vspace{0.3cm}

Now we turn our attention to the expected gain $u_2$ of the contender {\bf 2} as a function of $\alpha_2$ for fixed values of its first argument $\alpha_1=t$.

\begin{equation}
\label{eq3.14}
u_2(t,\alpha_2) = (t + \alpha_2 r) \frac{(\alpha_2 r)^{\gamma_r}}{(\alpha_2 r)^{\gamma_r}+t^{\gamma_r}} + \left( 1-t + (1-\alpha_2)r \right) \frac{((1-\alpha_2) r)^{\gamma_p}}{((1-\alpha_2) r)^{\gamma_p}+ (1-t)^{\gamma_p}} \;.
\end{equation}

\vspace{0.3cm}

For $t=0$ we have 
\begin{equation} 
\label{eq3.15}
u_2(0, \alpha_2) = \alpha_2 r + (1+(1-\alpha_2) r) \frac{((1-\alpha_2) r)^{\gamma_p}}{((1-\alpha_2) r)^{\gamma_p}+ 1} \;.
\end{equation}
This is a continuous function at the origin, 
\begin{equation} 
\label{eq3.16}
u_2(0,0^+) = u_2(0,0) =  \frac{(1+r) r^{\gamma_p}}{1+ r^{\gamma_p}} > r \;,
\end{equation}
and
\begin{equation} 
\label{eq3.17}
u_2(0,1) = r \;,
\end{equation}
then it is assured that $\beta_2(0) < 1$. The derivative of $u_2(0, \alpha_2)$ is easily calculated as
\begin{equation} 
\label{eq3.18}
u_2'(0,\alpha_2) = \frac{r}{1+((1-\alpha_2)r)^{\gamma_p}} \left( 1 - (1+(1-\alpha_2)r) \frac{\gamma_p ((1-\alpha_2)r)^{\gamma_p-1}}{1+((1-\alpha_2)r)^{\gamma_p}} \right) \;;
\end{equation}
it diverges to $- \infty$ at $\alpha_2=1$, and takes the value, at the origin,
\begin{equation} 
\label{eq3.19}
u_2'(0,0^+) = \frac{r}{1+r^{\gamma_p}} \left( 1 - \frac{(1+r) \gamma_p r^{\gamma_p-1}}{1+r^{\gamma_p}} \right) \;.
\end{equation}

\vspace{0.3cm}

After the change of variable $z=(1-\alpha_2)r$, the equation $u_2'(0,\alpha_2) = 0$ can be re-written as
\begin{equation} 
\label{eq3.20}
\gamma_p (1+z) = z + z^{\gamma_p - 1} \;,
\end{equation}
and note that, for $\alpha_1=0$, the variable $z$ is just the ratio of resources invested by the contenders in the poor-rewarding front:  $z=[(1-\alpha_2)/(1-\alpha_1)]r=(1-\alpha_2)y/(1-\alpha_1)x$.

\vspace{0.3cm}

An argument similar to the one used above with the equation (\ref{eq3.10}) convinces oneself that the equation (\ref{eq3.20}) has a unique positive solution $z^*(\gamma_p)$, that depends solely on the Tullock parameter $\gamma_p$, and it is a monotone increasing function of this parameter. As a consequence, the value of $z^*$ is bounded above by $z^*(1^-) \simeq 0.278465$, after noticing that the correct limit  when $\gamma_p \rightarrow 1$ of the equation (\ref{eq3.20}) is $1+z=- \ln z$.

\vspace{0.3cm}

Thus, provided the condition $r>z^*(\gamma_p)$ holds, the solution of equation $u_2'(0,\alpha_2) = 0$ is 
\begin{equation} 
\label{eq3.21}
\alpha_2^*(r, \gamma_p) = 1- \frac{z^*(\gamma_p)}{r} \;,
\end{equation}

\vspace{0.3cm}

\begin{figure}
  \centering
  \includegraphics[width=0.75\linewidth]{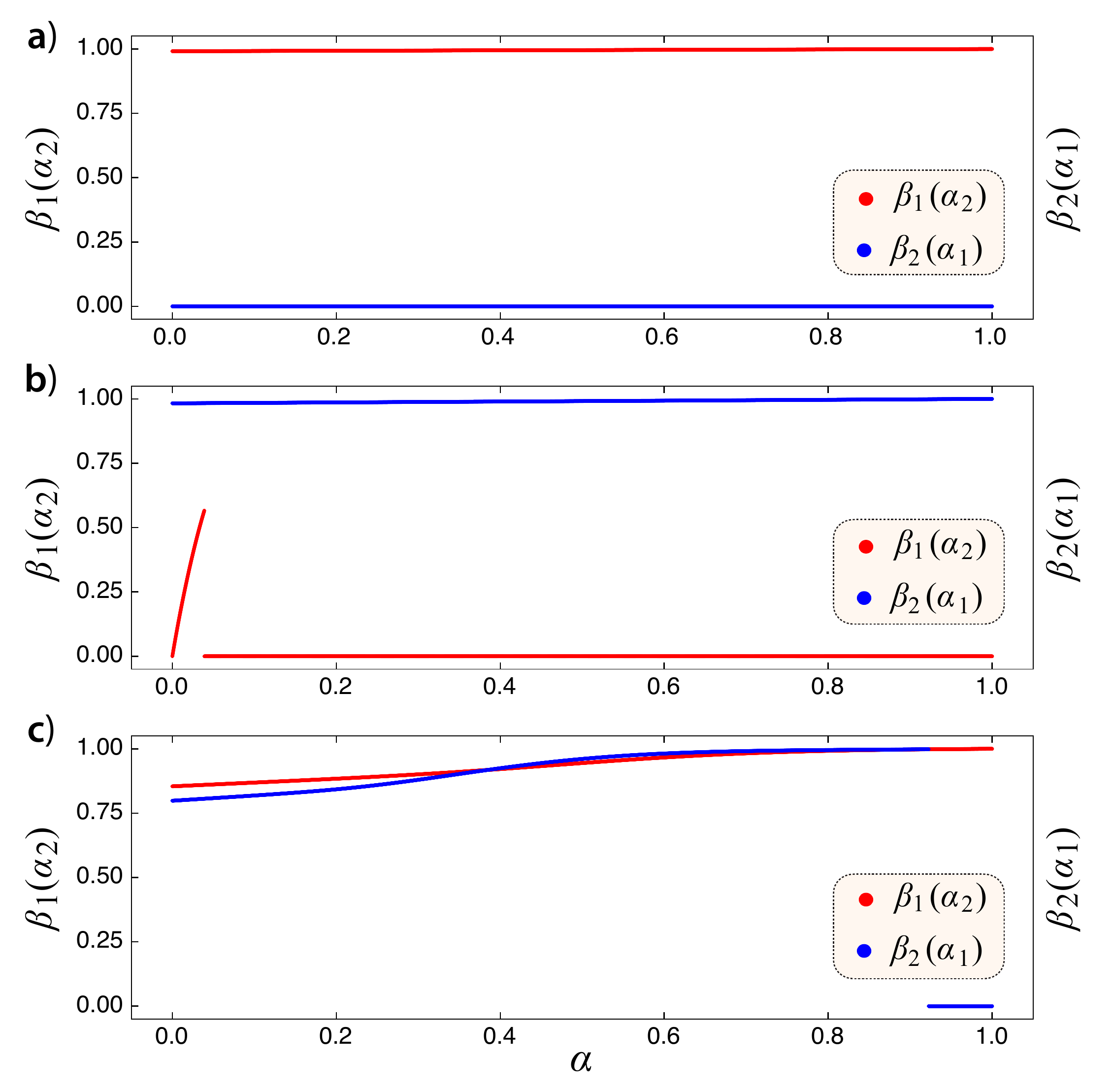}
  \caption{RR game with parameters $\gamma_r=5$, and $\gamma_p=0.5$. Plots of the best-response map $\beta_1(\alpha_2)$ of the contender {\bf 1} (red), together with the best-response map $\beta_2(\alpha_1)$ of the contender {\bf 2} (blue), for $r=0.05$ (Panel \textbf{a}), $r=0.1$ (\textbf{b}) and $r=0.85$ (\textbf{c}). In the regime of very small values of $r$ (Panel \textbf{a}, $r=0.05$), $\beta_2(\alpha_1)=0$ for all values of $\alpha_1$. In the intermediate regime of not too small values of $r< z^*(\gamma_p) \simeq 0.1715$ (Panel \textbf{b}, $r=0.1$) the map $\beta_2(\alpha_1)$ increases from zero with a relatively large slope before falling discontinuously to zero value. In the regime of $r> z^*(\gamma_p)$ (Panel \textbf{c}, $r=0.85$), $\beta_2(\alpha_1)$ increases from its non zero value $\alpha_2^*(r, \gamma_p)$, (see equation \ref{eq3.21}) at the origin and finally falls to zero.}
  \label{fig6}
\end{figure}

We are led to the conclusion that for values of $r< z^*(\gamma_p)$ the function $u_2(0, \alpha_2)$ is a monotone decreasing function of $\alpha_2$ and then $\beta_2(0) = 0$, while for $r>z^*(\gamma_p)$ the best-response to $t=0$ is $\beta_2(0) = \alpha_2^*(r, \gamma_p)$, the location of the local maximum of $u_2(0, \alpha_2)$, given by the equation (\ref{eq3.21}), which increases continuously from zero at $r = z^*(\gamma_p)$ up to the value $1-z^*$ at $r=1$.

\vspace{0.3cm}

For very small values of $t$ and $\alpha_2 \gg t$, $u_2(t, \alpha_2)$ is essentially given by $u_2(0, \alpha_2)$. However, for small values of $\alpha_2$ both functions are quite different. To see this, consider the partial derivative of $u_2(t, \alpha_2)$ respect to $\alpha_2$:
\begin{eqnarray}
u_2'(t, \alpha_2) & = & \frac{r (\alpha_2 r)^{\gamma_r}}{(\alpha_2 r)^{\gamma_r}+t^{\gamma_r}} +(t+\alpha_2 r) \frac{r \gamma_r (\alpha_2 r)^{\gamma_r -1} t^{\gamma_r}}{((\alpha_2 r)^{\gamma_r}+t^{\gamma_r})^2} \nonumber \\
 &  & \mbox{ } - \frac{r ((1-\alpha_2)r)^{\gamma_p}}{((1-\alpha_2)r)^{\gamma_p}+ (1-t)^{\gamma_p}} - (1-t+(1-\alpha_2)r) \frac{r \gamma_p ((1-\alpha_2 ) r)^{\gamma_p -1}(1-t)^{\gamma_p}}{(((1-\alpha_2)r)^{\gamma_p}+ (1-t)^{\gamma_p})^2} \;.
\end{eqnarray}
which takes the value, at $\alpha_2=0$,
\begin{equation} 
\label{eq3.22}
u_2'(t,0) = - \frac{r^{\gamma_p}}{r^{\gamma_p} + (1-t)^{\gamma_p}} \left(r + (1-t+r) \frac{\gamma_p (1-t)^{\gamma_p}}{r^{\gamma_p}+(1-t)^{\gamma_p}} \right) \;.
\end{equation}
We then see that, unlike $u_2'(0,0^+)$ whose sign depends on the $r$ value,  its limit when $t \rightarrow 0$ is negative, for all values of $r$:
\begin{equation} 
\label{eq3.23}
u_2'(0^+,0) = - \frac{r^{\gamma_p}}{1 + r^{\gamma_p} } \left(r + \frac{\gamma_p (1+r)}{1 +.r^{\gamma_p}} \right) < 0 \;,
\end{equation}
and, furthermore, from (\ref{eq3.19}) we have $u_2'(0^+,0) < u_2'(0,0^+)$ for all values of $r$. Thus the function $u_2(t, \alpha_2)$ decreases initially faster than $u_2(0, \alpha_2)$. Also, it is initially convex, before changing to concave in an interval of $\alpha_2$ values of the scale of $t/r$. Depending on the values of $r$, one observes three different behaviors for the location of its maximum, $\beta_2(t)$, for very small values of $t$: In the regime of very small values of $r$, $\beta_2(t) =0$. For values of $r> z^*(\gamma_p)$, $\beta_2(t)$ increases slowly from its value $\alpha_2^*(r, \gamma_p)$. In an intermediate regime of not too small values of $r< z^*(\gamma_p)$, the map $\beta_2(t)$ increases from zero with a relatively large slope.

\vspace{0.3cm}

For larger values of $t$, the best-response map $\beta_2(t)$ drops to a zero value in the last two regimes, while remaining at zero in the first regime. In other words, in the RR game, the best response of the contender {\bf 2} to any not-too-small (compared to $r$) investment of its (richer) opponent in the rich-rewarding front is to invest all of its resources in the poor-rewarding front. 

To illustrate these findings, in Figure \ref{fig6}, we show the graphs corresponding to the best-response maps for both contenders. Red dots display the maps for Contender \textbf{1} and blue ones for Contender \textbf{2}. Panels \textbf{a} (top), \textbf{b} (center), and \textbf{c} (bottom) correspond to  $r=0.05$, $r=0.1$, and $r=0.85$, respectively. On the one hand, the red dots show the predicted smooth increase of $\beta_1$ with $\alpha_2$. On the other hand, the blue dots display the three regimes predicted for $\beta_2(\alpha_1)$: i) for very small values of $r$ (here, $r=0.05$), $\beta_2(\alpha_1)=0$ for any $\alpha_1$; ii) for intermediate values of $r$ ($r< z^*(\gamma_p)\simeq 0.1715$, here $r=0.1$),  $\beta_2(\alpha_1)$ shows an increase from zero, through a steep slope, and then, through a discontinuity, goes to zero; iii) finally, for large values of $r$ ($r> z^*(\gamma_p)$, here $r=0.85$), $\beta_2(\alpha_1)$ increases from a strictly positive value $\alpha_2^*(r, \gamma_p)$ for $\alpha_1=0$, according to equation (\ref{eq3.21}), and finally, through a discontinuity, goes to zero.

\begin{figure}
  \centering
  \includegraphics[width=0.75\linewidth]{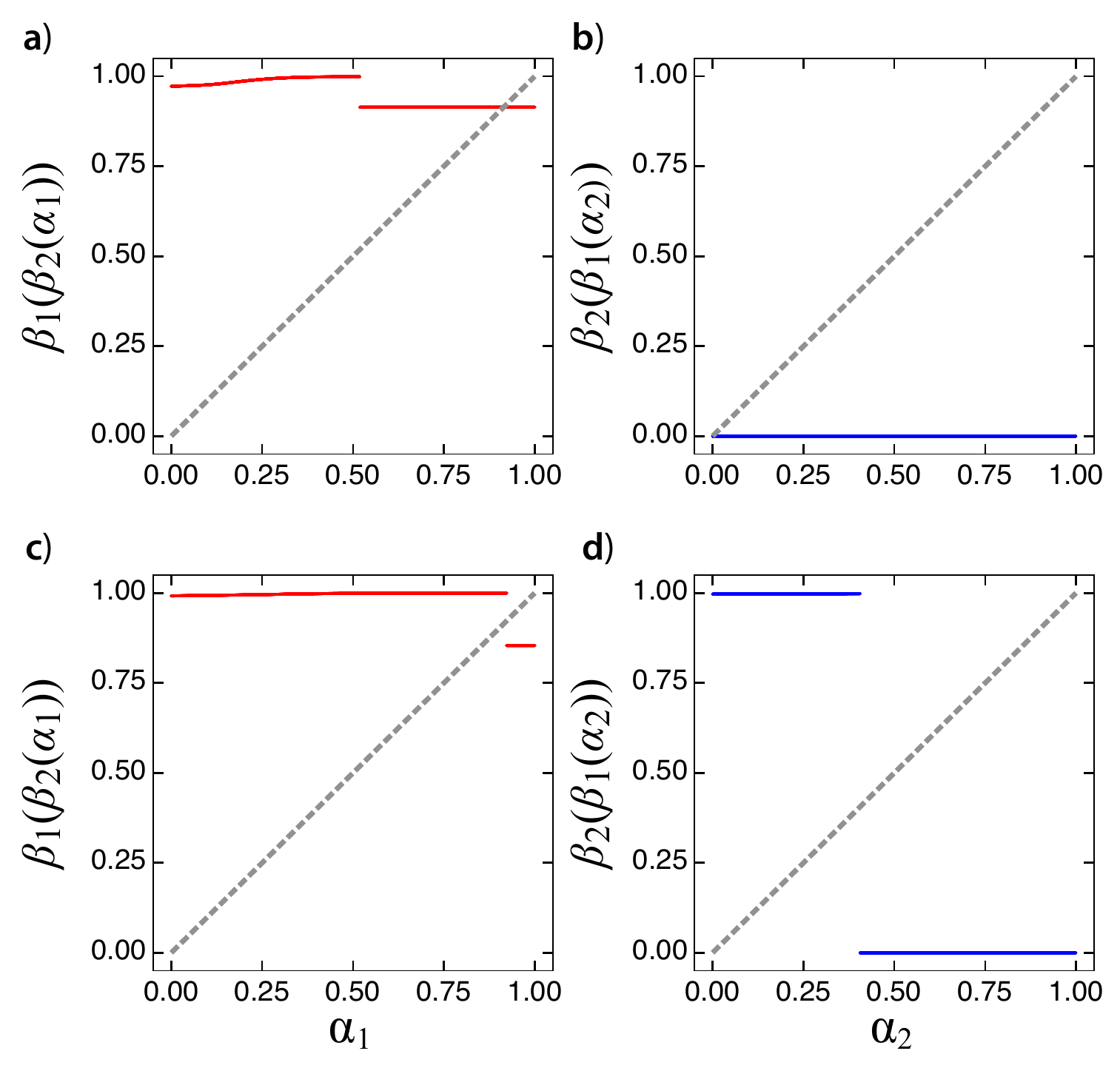}
  \caption{
RR game with $\gamma_r=5$ and $\gamma_p=0.5$. The top panels (\textbf{a} and \textbf{b}) display the composition of players' best-response maps for \textbf{$r=0.5$}, and the bottom panels \textbf{(c} and \textbf{d}) for \textbf{$r=0.85$}. Correspondingly, left panels (\textbf{a} and \textbf{c}) show $\beta_1(\beta_2( \alpha_1))$, while $\beta_2(\beta_1( \alpha_2))$ is shown in right panels (\textbf{b} and \textbf{d}). The main diagonal (in dashed black) is plotted to visualize the existence of Nash equilibrium for $r=0.5$, and its absence for $r=0.85$.
}
  \label{fig7}
\end{figure}



\vspace{0.3cm}

\subsection{The Nash equilibrium}
\label{subsec:RR_ne}

The analysis of the best-response map $\beta_2(t)$ of the contender {\bf 2} indicates its marked overall preference for investing all its resources in the poor-rewarding front. On the other hand, we have also shown that the best-response of the contender {\bf 1} to that eventuality is $\beta_1(0)=1-\frac{r}{z^*}$, equation (\ref{eq3.11}). Consequently, if it is the case that
\begin{equation}
\label{eq3.24}
\beta_2\left(1-\frac{r}{z^*}\right)=0 \;,
\end{equation}
we are led to the conclusion that the pair $(1-\frac{r}{z^*},0)$ is a Nash equilibrium of the RR game. Let us remark here that $z^*(\gamma_p)$ is bounded below by $z^*(1^-) \simeq 3.590175 > 1$, so that $1-\frac{r}{z^*}$ is bounded below by $0.721462$, not a small quantity.

\vspace{0.3cm}

The expected gain $u_2(t, \alpha_2)$, at $t=1-\frac{r}{z^*}$, are given by

\begin{equation}
\label{eq3.25}
u_2 \left(1-\frac{r}{z^*}\,, \alpha_2  \right) = \left( 1-\frac{r}{z^*} + \alpha_2 r \right) \, \frac{(\alpha_2 r)^{\gamma_r}}{(1 - r/z^*)^{\gamma_r} + (\alpha_2 r)^{\gamma_r}} + \frac{r}{z^*} ( 1 + z^*(1 - \alpha_2))  \frac{(z^*(1 - \alpha_2))^{\gamma_p}}{1 +(z^*(1 - \alpha_2))^{\gamma_p} } \;.
\end{equation}

For small values of $r$, the dominant term in (\ref{eq3.25}) is the second term (linear in $r$) in the RHS, because $\gamma_r > 1$. This term is maximum at $\alpha_2=0$, and this proves that at least for small values of $r$, one has $\beta_2(1-\frac{r}{z^*}) =0$.

\vspace{0.3cm}

\begin{figure}
  \centering
  \includegraphics[width=0.75\linewidth]{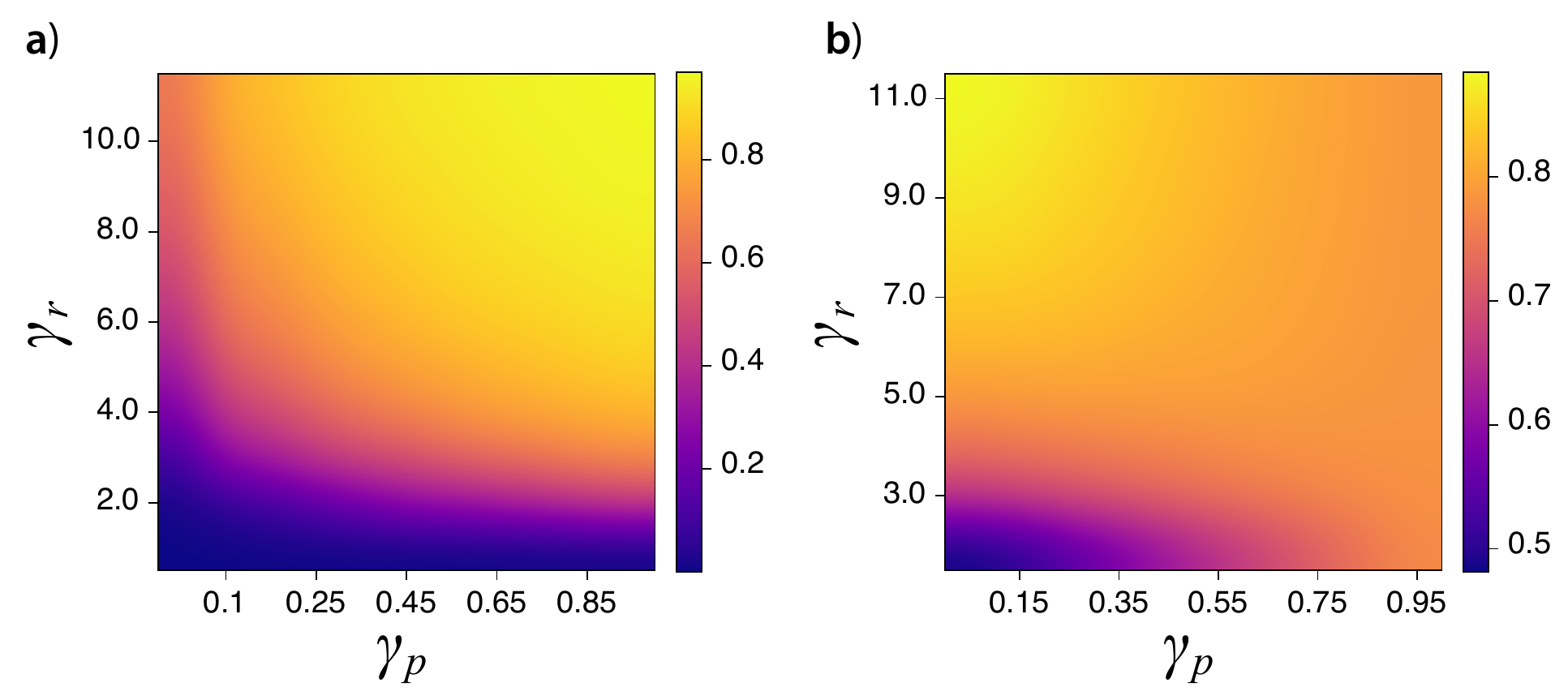}
  \caption{Heat maps showing the value of the threshold value $r_{th}$ for KR game (Panel \textbf{a}) and RR game (Panel \textbf{b}) in the space $(\gamma_r, \gamma_p)$. Results have been obtained through numerical exploration. Recall that in the KR game, only when $r<r_{th}^{KR}$ the Nash equilibrium disappears and the contenders have no incentive to fight, whereas in the RR game it is when $r>r_{th}^{RR}$ that peace sets in.}
  \label{fig8}
\end{figure}

To exemplify these findings, Figure \ref{fig7} displays the best-response maps corresponding to the Tullock parameters $\gamma_r=5$ and $\gamma_p=0.5$ for $r=0.5$ (panels \textbf{a} and \textbf{b}) and $r=0.85$ (\textbf{c} and \textbf{d}). Left panels (\textbf{a} and \textbf{c}) show the $\beta_1(\beta_2( \alpha_1))$ maps and right ones (\textbf{b} and \textbf{d}) the $\beta_2(\beta_1( \alpha_2))$ ones. Nash equilibria would be denoted by an inner intersection of the curve with the black main diagonal. Our extensive numerical exploration in the parameter plane $(\gamma_r, \gamma_p)$ indicates the existence of an upper bound $r_{th}(\gamma_r, \gamma_p)$ such that if $r < r_{th}^{RR}$, the equation (\ref{eq3.24}) holds, and then the pair $(1-\frac{r}{z^*},0)$ is a Nash equilibrium of the RR game. As a numerical example, for $\gamma_r=5$ and $\gamma_p=0.5$, we find the value $r_{th}^{RR} = 0.790541$. Figure \ref{fig8} depicts the landscape of threshold values $r_{th}$ for both games, KR (Panel \textbf{a}) and RR (Panel \textbf{b}).

\vspace{0.3cm}

\vspace{0.3cm}

\section{Repeated combat}
\label{sec:repeated}

\vspace{0.3cm}

Let us note that, as we have shown above in section \ref{sec:KR}, the expected gain of contender {\bf 1} in the KR game $u_1(\bar{\alpha}, \bar{\alpha}) > 1$ is greater than its initial resources, and then, whenever the Nash equilibrium exists for the KR game there is an incentive for him/her to fight. In a similar way, in section  \ref{sec:RR} we have seen that provided a Nash equilibrium exists for a RR game, the contender {\bf 1} earns $1-\frac{r}{z^*}$ with certainty in the rich-rewarding front, and also that, as its investment in the poor-rewarding front, $\frac{r}{z^*} < r$ is lower than its opponent investment, its expected gain in this front is larger than its investment, and then there is an incentive for the contender {\bf 1} to fight in a RR game.

\vspace{0.3cm}

As a consequence, for both KR and RR games, it seems rather natural to assume that in the eventuality that combat ends in a tie, the combat will be repeated, until either {\it a}) one of the contenders reaches a victory in both fronts or, {\it b}) as it may happen in the RR game where resources are redistributed when tying, a Nash equilibrium no longer exists after the tie.

\vspace{0.3cm}

First, we analyze in subsection \ref{subsec:repeated_KR} the repeated KR game, where we will reach a somewhat surprising simple result, namely that the repeated KR game is equivalent to a non-repeated game in one front with a Tullock CSF with a parameter that is the sum of those of the CSF fronts' functions, $\gamma_r$ and $\gamma_p$. In subsection \ref{subsec:repeated_RR} we study the repeated RR game and show that under the condition that a game is played, i.e. combat takes place, only if a Nash equilibrium exists (and contrary to the KR game), it is not equivalent to a single non-repeated game in one front, for there is a non-zero probability of reaching a situation in which a Nash equilibrium does not exist. 

\vspace{0.3cm}

\begin{figure}
  \centering
  \includegraphics[width=0.75\linewidth]{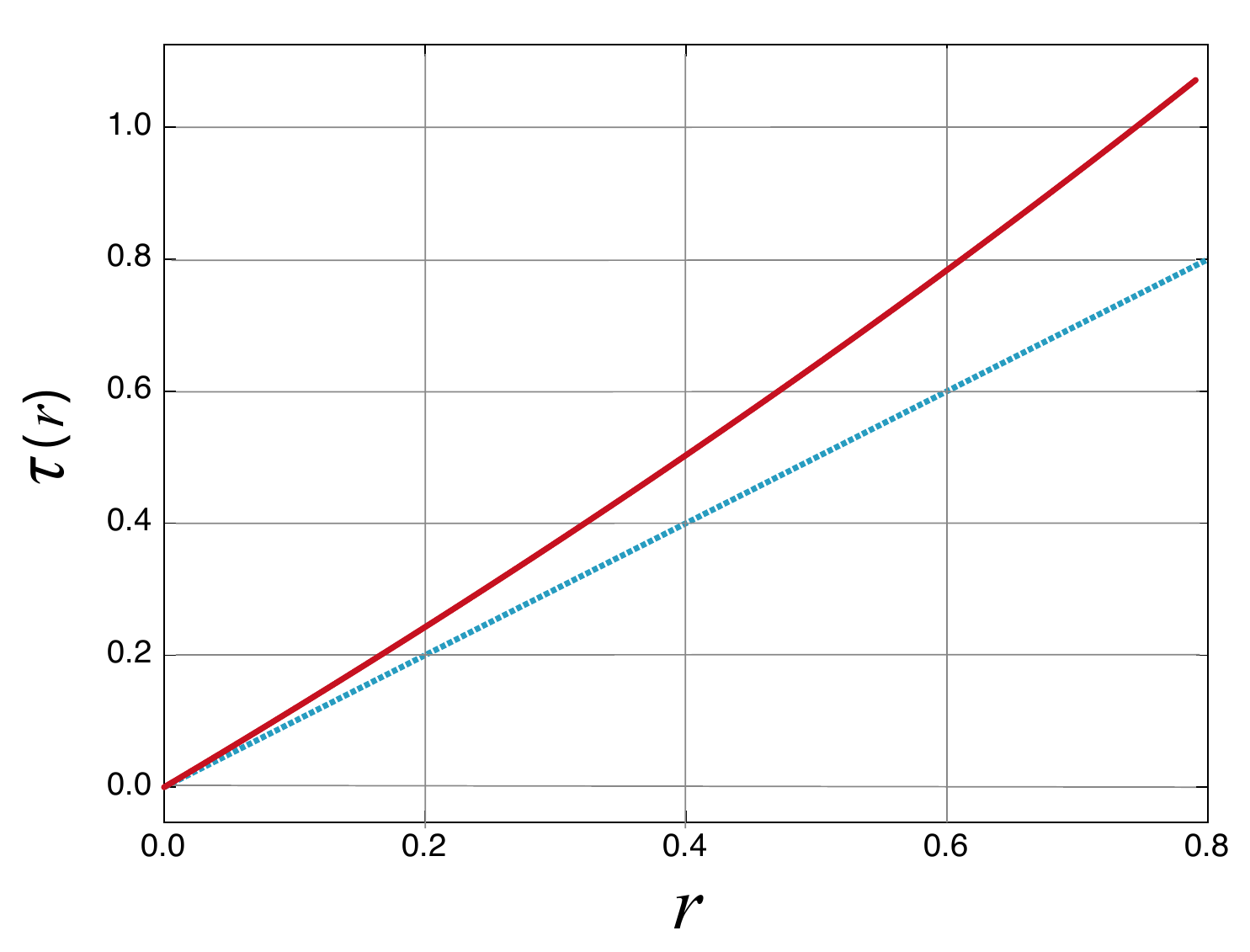}
  \caption{Map $\tau(r)$ representing the rescaled resources of contender {\bf{2}} after a repeated RR game, thus a tie, departing from a resource base of $r$ in the initial game. The dashed diagonal line $\tau(r)=r$ marks the boundary where resources after a tie would be the same as before.}
  \label{fig9}
\end{figure}

\subsection{Repeated KR game}
\label{subsec:repeated_KR}

Assuming a Nash equilibrium ($\bar{\alpha}, \bar{\alpha}$) of the KR game exists, see equations (\ref{eq2.13}) and (\ref{eq2.14}), let us simply denote by $\bar{p}$ (resp. $\bar{q}$) the probability of victory, at the Nash equilibrium values of investments, of the contender {\bf 1} in the rich-rewarding (resp. poor-rewarding) front, i.e.
\begin{equation}
\label{eq2.17}
\bar{p} = (1 + r^{\gamma_r})^{-1} \;, \;\;{\mbox{and}}\;\;\; \bar{q} = (1 + r^{\gamma_p})^{-1}\;,
\end{equation}
so that the probability of a tie is $\bar{p}(1- \bar{q}) + \bar{q}(1- \bar{p}) = \bar{p} + \bar{q} - 2\bar{p} \bar{q}$.

\vspace{0.3cm}

In a KR game, the situation after an eventual tie is just the initial one, and these probabilities are thus unchanged. Now, the probability $p_{\infty}$ that the repeated combats end in a victory of contender {\bf 1} is 
\begin{equation}
\label{eq2.18}
p_{\infty} = \sum_{k=1}^{\infty} (\bar{p} + \bar{q} - 2\bar{p} \bar{q})^k \bar{p} \bar{q} = \frac{ \bar{p} \bar{q}}{(1 -\bar{p} - \bar{q} + 2\bar{p} \bar{q})} = \frac{1}{1+r^{(\gamma_r + \gamma_p)}}\;.
\end{equation}

This, somehow unexpectedly simple, result can be stated in the following way: Provided a Nash equilibrium exists for a KR game with Tullock parameters $\gamma_r$ and $\gamma_p$, the repeated game is equivalent to a single combat with a Tullock parameter $\gamma_r + \gamma_p$, i.e. a single combat with a CSF that is more rich-rewarding than any of the original ones. Indeed, after a second thought, given that the incentive to fight a single combat is on the rich contender's side, the result shouldn't come as much surprise, for the repetition of it can only increase the (cumulative) expected gain. Nonetheless, we find it remarkable that the set of Tullock functions is, in this particular (and admittedly loose, in need of precision) sense, a closed set under the ``repetition operation''.

\vspace{0.3cm}

\subsection{Repeated RR game}
\label{subsec:repeated_RR}

Assuming that a Nash equilibrium exists for a RR game, a tie occurs whenever the contender {\bf 2} reaches victory in the poor-rewarding front. Thus the probability of a tie in a single combat is 
\begin{equation}
\label{eq2.18}
p_t = \frac{(z^*)^{\gamma_p}}{1 +(z^*)^{\gamma_p}}\;,
\end{equation}
where it should be noted that (as $z^*>1$)  $p_t>1/2$. In other words, a tie has a larger probability than a victory of the contender {\bf 1}. Also, note that
this probability is independent of the resources $r$ of the contender {\bf 2}. Consequently, though the resources of the contenders change after a tie, this probability remains unchanged, provided there is a Nash equilibrium after redistributing resources.

\vspace{0.3cm}

\begin{figure}
  \centering
  \includegraphics[width=0.75\linewidth]{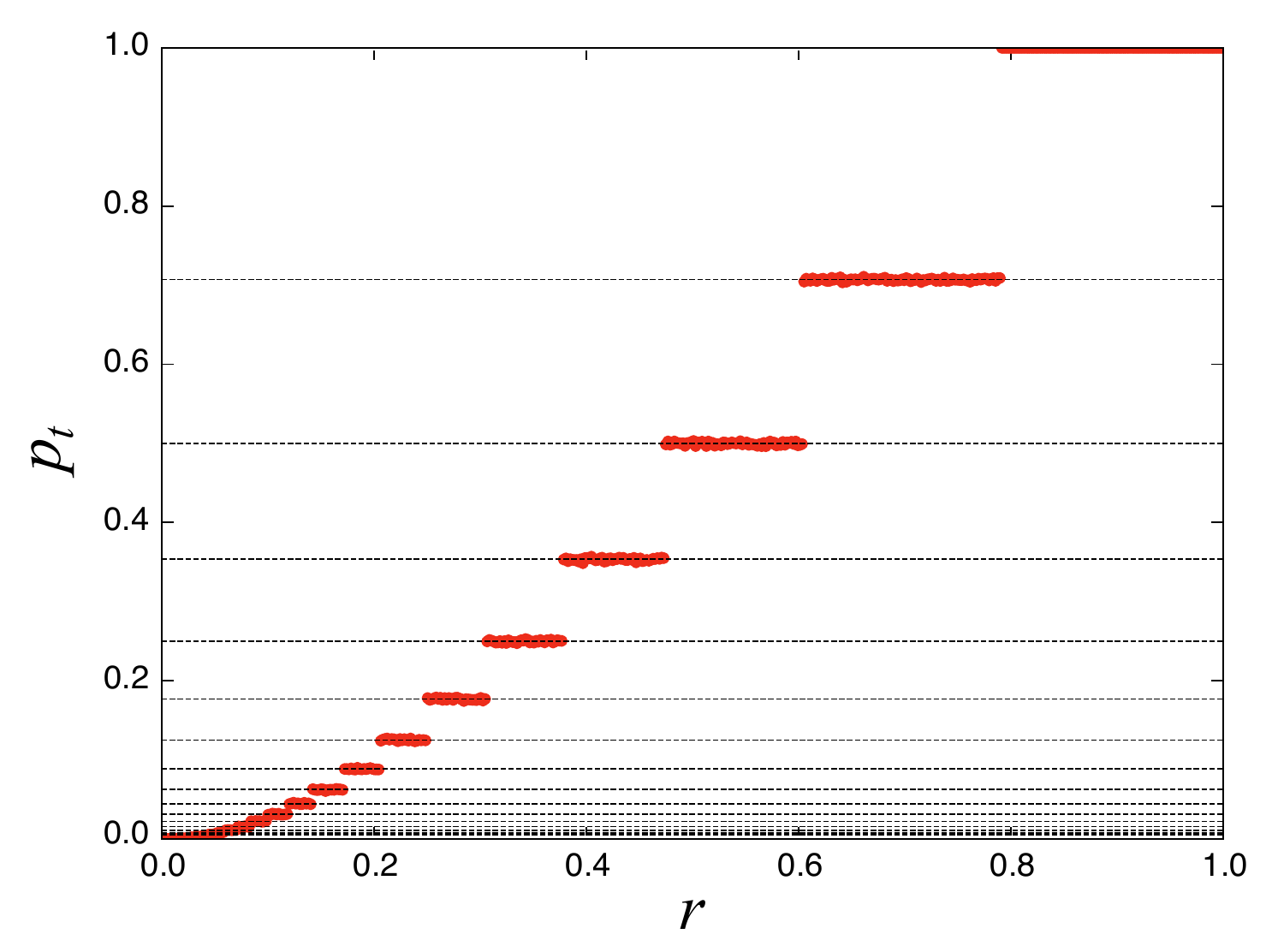}
  \caption{Probability of tying in RR game as a function of the resource ratio $r$ in the first round of the game. Every red dot represents the probability of a tie event in a Monte Carlo simulation averaged over $10^5$ realizations. Horizontal dashed lines represent the analytical result as given by $\rho(x)$. Theory and simulations match perfectly.}
  \label{fig10}
\end{figure}

\begin{figure}
  \centering
  \includegraphics[width=0.75\linewidth]{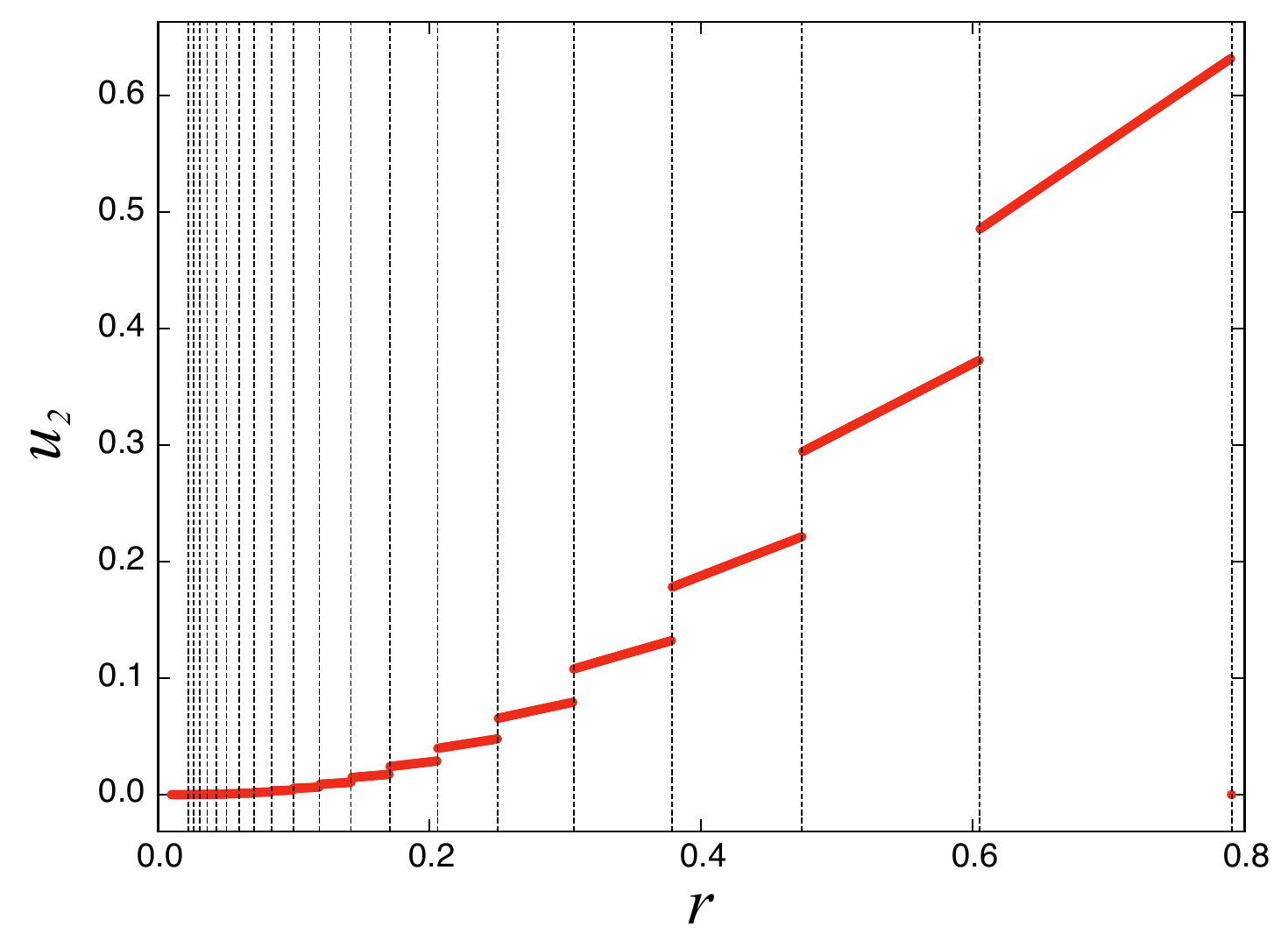}
  \caption{Expected gain $u_2^{\mathrm{rep}}$ of contender {\bf{2}} for the repeated RR game as a function of the resource ratio $r$ at the beginning of the game. Straight dashed vertical lines mark the succesive $n$ jumps performed with map $\tau^{-n}(r_{th}^{RR})$. Results shown for $\gamma_r=5$ and $\gamma_p=0.5$.}  
  \label{fig11}
\end{figure}

After a tie occurs, the contender {\bf 1} resources become $1-\frac{r}{z^*}$, while those of contender {\bf 2} are now $r(1+\frac{1}{z^*})$. For the analysis of the repeated RR game, it is convenient to rescale the new resources of the contenders, so that the rescaled resources are 1 for the contender {\bf 1} and 
\begin{equation}
\label{eq2.19}
\tau(r) = \frac{r(1+z^*)}{z^*-r}  
\end{equation}
for the contender {\bf 2}. The map defined by equation (\ref{eq2.19}) is a continuous monotone (thus invertible) increasing map with a slope larger than 1 for all $r$. Figure \ref{fig9} depicts the graph of this map.

\vspace{0.3cm}

If it is the case that $\tau(r)<r_{th}^{RR}$, a Nash equilibrium for the ``rescaled'' RR game exists, and then the contender {\bf 1}, despite its recent defeat and the fact that she owns lower resources than before, has the incentive to fight and thus the combat is repeated. Otherwise, if $r_{th}^{RR}<\tau(r)<(r_{th}^{RR})^{-1}$, there is no Nash equilibrium after the tie, and none of the contenders has the incentive to fight. The eventuality that $\tau(r) > (r_{th}^{RR})^{-1}$ (being $r < r_{th}^{RR}$) would require $r_{th}^{RR} > z^*/(1+z^*)$, a condition that we have never found in our extensive numerical exploration of the $r_{th}^{RR}$ values in the plane $(\gamma_r, \gamma_p)$. This observation excludes the possibility that a repeated RR game could end in a final victory of contender {\bf 2}. Incidentally, there are situations where for some interval of values of $r<r_{th}^{RR}$, $\tau(r)>1$; in these cases, the repeated RR game ends with interchanged (rich-poor) contenders' role.

\vspace{0.3cm}

We have been led to the conclusion that there are two mutually exclusive outcomes for a repeated RR game, namely either a victory of the contender {\bf 1} or a final situation of survival of the two contenders with no Nash equilibrium, that we will briefly call peace. For fixed values of $\gamma_r$ and $\gamma_p$, we define the function $\rho(r)$ (for $r$ in the open interval $(0,1)$) as the probability that a repeated RR game where the ratio of the resources is $r$ ends in peace. This function can be computed once the values of $z^*(\gamma_p)$ and $r_{th}^{RR}(\gamma_r, \gamma_p)$ have been numerically determined.

\vspace{0.3cm}

The function $\rho(r)$ is a piecewise constant, i.e. a staircase. It takes the value 1 for $r_{th}^{RR}<r<1$. If $\tau^{-1}(r_{th}^{RR}) < r < r_{th}^{RR}$, a tie occurs with probability $p_t$, after which peace is reached, so $\rho(r)= p_t$ for $r$ in this interval, and so on. Then
\[ \rho(x) = \left\{ \begin{array}{ll}
1 & \mbox{if $r_{th}^{RR}<r<1$} \\
p_t^n & \mbox{if $\tau^{-n}(r_{th}^{RR}) < r < \tau^{-n+1} (r_{th}^{RR})$, $n=1, 2, ...$}
\end{array} \right. \]

In Figure \ref{fig10} it is shown how this expression matches the stochastic simulations performed on the RR game with $\gamma_r=5$ and $\gamma_p=0.5$.

\vspace{0.3cm}

The computation of the expected gain $u_2^{{\text{rep}}}(r)$ of contender {\bf 2} for the repeated RR game requires undoing the rescaling of resources made at each iteration of the map $\tau$. The rescaling factor for the $i$-th iteration is $1- \tau^{i-1}(r)/z^*$, and thus if $\tau^{-n}(r_{th}^{RR}) < r < \tau^{-n+1} (r_{th}^{RR})$, after $n$ repeated tying contests ending in a peaceful situation, the final resources of contender {\bf 1} will be $$ \Pi_{i=1}^n \left(1-\frac{\tau^{i-1}(r)}{z^*}\right) \;, $$ and then 

\begin{equation}
 u_2^{{\text{rep}}}(r) =
p_t^n \left(1 + r - \Pi_{i=1}^n \left( 1- \frac{\tau^{i-1}(r)}{z^*} \right)\right) \;\; \mbox{if $\tau^{-n}(r_{th}^{RR}) < r < \tau^{-n+1} (r_{th}^{RR})$, $n=1, 2, ...$}\;.
\end{equation}
Figure \ref{fig11} shows the staircase form for $u_2^{\text{rep}}$ together with the boundaries marked by the inverse map $\tau^{-n}(r_{th}^{RR})$, $n=1,2,...$ Computations have been done, as usual, for $\gamma_r=5$ and $\gamma_p=0.5$.

\vspace{0.3cm}

\section{Concluding remarks}
\label{sec:conclusion}

In this work, we have explored the resolution of conflicts under the probabilistic framework of Tullock's combat success functions and game theory. These functions depend on the ratio resources of the contenders, $r=y/x$, and a parameter $\gamma$, called the technology parameter. In particular, we have focused on conflicts taking part simultaneously on two fronts. Each front is characterized by a different value of $\gamma$, being one front rich-rewarding ($\gamma_R>1$), where the richer contender has incentives to fight, and the other poor-rewarding ($0<\gamma_P<1$), where the poorer may take the lead. We define the game or combat in such a way that if a contender wins on both fronts, takes all of the adversary resources plus their initial resources, $1+r$, and the one losing is defeated and the game is over. Not all resolutions lead to a total victory, if a contender wins one front but loses the other, a tie happens. In case of a tie, different scenarios are possible in order to reward/punish the contenders and allow for the next round. Here, we proposed two and thus gave birth to two different games as a consequence. In the keeping resources (KR) game, after a tie, both contenders conserve their original resources and simply a next round takes place. In the redistributing resources (RR) game, the winner of each front gains all the resources deployed at that front. These different rules give rise to different conflict dynamics and resolutions. 

\vspace{0.3cm}

Just by performing elementary mathematical analysis on the expected gain functions and the best-response maps for each player we can almost fully characterize each game. However, in order to gain a full understanding of the situation, the analytical results were checked and extended with numerical analysis and extensive simulations of the conflict dynamics. 

\vspace{0.3cm}

We remark the following main results. For both games there exists a threshold value of the resource ratio $r$ separating a regime where a Nash equilibrium exists in the best-response dynamics between contenders and a regime where this does not happen. This threshold is solely determined by the tuple of Tullock technology parameters $(\gamma_R, \gamma_P)$. In case of the existence of that equilibrium, combat takes place whereas if not, the contenders remain at peace. In the KR game, it is found that the peaceful regime occurs for $r \in [0, r_{th}^{KR})$, whereas in the RR game this happens for $r\in (r_{th}^{RR}, 1)$. 

\vspace{0.3cm}

In particular, for the KR game, when a Nash equilibrium exists, it is found that the investment fractions maximizing the contenders' expected gains are identical, $\bar{\alpha}$. The existence of a Nash equilibrium is subjected to a set of conditions. The value of $\bar{\alpha}$ must be a global maximum for both expected gains and it turns out that for a certain set of values of $r$, this condition does not always hold.  It is also found that provided a Nash equilibrium exists with Tullock parameters $\gamma_R$ and $\gamma_P$ for the KR game, the repeated game is equivalent to a game with a single front where the technology parameter is $\gamma_R+\gamma_P$, the sum of the parameters at both fronts and thus it is equivalent to a more rich-rewarding front. 

\vspace{0.3cm}

In the RR game, when a Nash equilibrium exists, it is found that the investment fraction at the rich-rewarding front for Contender \textbf{2} is always zero, while for Contender \textbf{1} an analytical expression is found (not holding in the peaceful regime, indeed). In this game, a tie occurs whenever Contender \textbf{2} reaches victory in the poor-rewarding front.  As resources are redistributed after a tie, the repeated RR game involves a richer behavior than the KR game. It is found that, provided a Nash equilibrium exists, the tie outcome occurs with a probability $p_t$ higher than the total victory of Contender \textbf{1} (winning at both fronts) and this probability is independent of $r$ and ultimately determined by $\gamma_P$. Redistribution after a tie always leads to the enrichment of the poorer contender and impoverishment of the richer one and thus the resource ratio after every round tends to increase. If repetition continues, eventually $r>r_{th}^{RR}$, and thus there is no incentive to fight for any contender. While there is a possibility to surpass $r>1$ from a higher enough $r<r_{th}^{RR}$, these jumps cannot overcome $r=1/r_{th}^{RR}$, and thus nonexistence of a Nash equilibrium still holds. We conclude for this repeated game that there are two mutually exclusive outcomes, namely either a victory of Contender \textbf{1} or a final situation of survival of the two contenders with no Nash equilibrium, a state of peace, where the contenders' resource difference has diminished. This repeated RR game dynamic is nicely represented in the staircase diagram, formulated analytically and perfectly reproduced by simulations, that depicts the probability of reaching a tie as a function of the resource ratio $r$ in the first round.

\vspace{0.3cm}

Throughout this analysis, we have assumed perfect rationality for the contenders involved together with perfect information. We recognize that these assumptions may be in general too rigid in order to translate our analysis and conclusions into practical applications. Thus, a direction of future work demands clearly a relaxation of some of these hypotheses. Another readily possible extension of the model could be to include more realism on the managing and deployment of resources by the contenders. Finally and most importantly, we have restricted ourselves to thoroughly analyzing the conflict involving just two agents and thus pairwise interactions. Of course, the reality is more complex and conflict may involve an arbitrarily large number of entities or contenders, each of it with its own particularities while interacting in complex ways (i.e. higher-order interactions). For this, the frameworks of complex networks and hypergraphs arise as very suggestive tools to extend this conflict dynamic to large heterogeneous systems.

\acknowledgments
A.dM.A. is funded by an FPI Predoctoral Fellowship of MINECO. We acknowledge partial support from the Government of Aragon, Spain, and “ERDF A way of making Europe” through grant E36-20R (FENOL) to A.dM.A, C.G.L, M.F. and Y. M., from Ministerio de Ciencia e Innovaci\'on, Agencia Espa\~nola de Investigaci\'on (MCIN/ AEI/10.13039/501100011033) Grant No. PID2020-115800GB-I00 to A.dM.A, C.G.L., M.F. and Y.M. 



 

\bibliographystyle{vancouver}
\bibliography{war.bib}




\end{document}